\documentclass[3p, twocolumn]{elsarticle}
\usepackage[colorlinks=true,pdfstartview=FitV,linkcolor=blue,citecolor=blue,urlcolor=blue,bookmarks=false]{hyperref}
\usepackage[usenames,dvipsnames]{color}
\usepackage[sc]{mathpazo}
\usepackage[below]{placeins}
\usepackage{afterpage}
\usepackage[separate-uncertainty,retain-explicit-plus,per-mode = symbol]{siunitx}
\usepackage{longtable}
\usepackage{multirow}
\usepackage{rotating}
\usepackage[displaymath, mathlines]{lineno}
\usepackage{graphicx}
\usepackage{paralist}
\usepackage{wrapfig}
\usepackage{appendix}
\usepackage{vruler}
\usepackage{etoolbox}
\usepackage{tikz}
\usepackage{listings}
\usepackage{amsmath}
\usepackage{subfigure}
\usepackage{relsize}
\usepackage{etoolbox}
\modulolinenumbers[5]
\newcolumntype{d}[1]{D{.}{\cdot}{#1}}
\let\oldding\ding
\renewcommand{\ding}[1]{\ifnum#1=73 $\star$\else\oldding{#1}\fi}

\DeclareSIUnit\c{$c$}
\DeclareSIUnit\week{w}
\DeclareSIUnit\year{yr}
\DeclareSIUnit\standard{std}
\DeclareSIUnit\str{sr}
\DeclareSIUnit\pe{PE}
\DeclareSIUnit\ev{events}
\DeclareSIUnit\bin{(5-PE bin)}
\DeclareSIUnit\hit{hits}
\DeclareSIUnit\sgm{$\sigma$}
\DeclareSIUnit\rms{RMS}
\DeclareSIUnit\keVr{keV$_{\rm r}$}
\DeclareSIUnit\keVee{keV$_{\rm $e$e}$}
\DeclareSIUnit\ph{photons}

\DeclareSIUnit\sample{samples}
\DeclareSIUnit\counts{counts}
\DeclareSIUnit\countsamples{count\mathord{\cdot}samples}
\DeclareSIUnit\pepertrigger{PE/trigger}

\newcommand{\pmt}{\mbox{PMT}}
\newcommand{\pmts}{\mbox{PMTs}}

\newcommand{\spe}{\mbox{SPE}}
\newcommand{\clt}{\mbox{CLT}}
\newcommand{\blankdata}{blank}
\newcommand{\laserdata}{laser}
\newcommand{\zerope}{zero-pe}
\newcommand{\nonzerope}{non-zero-pe}
\newcommand{\thresholdcut}{threshold}

\newcommand{\sect}{Section~}
\newcommand{\append}{}
\newcommand{\eqn}{Eq.~}
\newcommand{\eqns}{Eqs.~}

\newcommand{\totaldist}{\ensuremath{T(q)}}
\newcommand{\signaldist}{\ensuremath{S(q)}}
\newcommand{\signaldistv}[1]{\ensuremath{S_{#1}(q)}}
\newcommand{\bkgddist}{\ensuremath{B(q)}}
\newcommand{\pedist}{\ensuremath{L(p)}}
\newcommand{\pedistzero}{\ensuremath{L(0)}}
\newcommand{\pedistone}{\ensuremath{L(1)}}
\newcommand{\spedist}{\ensuremath{\psi(q)}}
\newcommand{\spedistconv}[1]{\ensuremath{\psi^{#1}(q)}}
\newcommand{\cumuldist}[2]{\ensuremath{F_{#1}(#2)}}
\newcommand{\totalrand}{\ensuremath{T}}
\newcommand{\signalrand}{\ensuremath{S}}
\newcommand{\bkgdrand}{\ensuremath{B}}
\newcommand{\sperand}{\ensuremath{\psi}}
\newcommand{\sperandconv}[1]{\ensuremath{\psi^{#1}}}
\newcommand{\perand}{\ensuremath{L}}
\newcommand{\occup}{\ensuremath{\lambda}}

\newcommand{\mean}[1]{\ensuremath{\boldsymbol{E} \left[#1\right]}}
\newcommand{\meansq}[1]{\ensuremath{\boldsymbol{E}^2\left[#1\right]}}
\newcommand{\var}[1]{\ensuremath{\boldsymbol{V}\left[#1\right]}}
\newcommand{\stdv}[1]{\ensuremath{\boldsymbol{SD}\left[#1\right]}}

\newcommand{\sampmean}[1]{\ensuremath{\boldsymbol{\widehat{E}}\left[#1\right]}}

\newcommand{\sampstdv}[1]{\ensuremath{\boldsymbol{\widehat{SD}}\left[#1\right]}}
\newcommand{\sampvar}[1]{\ensuremath{\boldsymbol{\widehat{V}}[#1]}}

\newcommand{\estim}[1]{\ensuremath{\widehat{#1}}}
\newcommand{\biasestim}[1]{\ensuremath{\widehat{#1}}}
\newcommand{\biassampmean}[1]{\ensuremath{\boldsymbol{\widehat{E}}\left[#1\right]}}

\newcommand{\nsamps}{\ensuremath{N}}
\newcommand{\nsampszero}{\ensuremath{N_0}}
\newcommand{\nsampsbkgd}{\ensuremath{N_B}}
\newcommand{\nlaseramp}{\ensuremath{A_T}}

\newcommand{\zeropefrac}{\ensuremath{f}}
\newcommand{\speleakfrac}{\ensuremath{k(\zeropefrac)}}
\newcommand{\leakmean}{\ensuremath{l}}
\newcommand{\simmean}[1]{\ensuremath{\boldsymbol{E}_{sim}\left[#1\right]}}
\newcommand{\simstdv}[1]{\ensuremath{\boldsymbol{SD}_{sim}\left[#1\right]}}
\newcommand{\simvar}[1]{\ensuremath{\boldsymbol{V}_{sim}\left[#1\right]}}
\newcommand{\betadist}{\text{Beta}}
\newcommand{\binomdist}{\text{Binom}}

\newcommand{\loge}[1]{\ensuremath{\ln\left(#1\right)}}

\newcommand{\conv}{\ensuremath{\ast}}

\newcommand{\approxim}{\ensuremath{\approx}}
\newcommand{\defined}{\ensuremath{\equiv}}

\newcommand{\pmtmodel}{R11410}
\newcommand{\dividerresistance}{\SI{37} {\mega\ohm}}
\newcommand{\terminatorresistance}{\SI{50} {\ohm}}
\newcommand{\pmtgainseventeen}{$1.6\times10^7$}
\newcommand{\pmtgainsixteen}{$9.8\times10^6$}
\newcommand{\pmtgainfifteen}{$5.9\times10^6$}
\newcommand{\pmtgainfourteen}{$3.5\times10^6$}

\newcommand{\laserpulsewidth}{\SI{60}{\pico \second}}     
\newcommand{\laserwavelength}{\SI{405}{nm}}   
\newcommand{\laserpulserate}{\SI{1}{\kilo \hertz}}
\newcommand{\laseracquiwindow}{\SI{1}{\micro\second}}

\newcommand{\ampgain}{$10$x}
\newcommand{\digitizerbit}{12 bit}
\newcommand{\digitizersamplerate}{\SI{250}{\mega \hertz}}
\newcommand{\digitizermodel}{V1720}
\newcommand{\triggerdelay}{\SI{500}{\nano\second}}

\newcommand{\baselinewindow}{\SI{20}{\nano\second}}
\newcommand{\laserintegrationwindow}{\SI{208}{\nano\second}}

\newcommand{\zeropefrachighthreshold}{\SI{0.333}{}}
\newcommand{\zeropefraclowthreshold}{\SI{0.1}{}}

\newcommand{\darknoisefraction}{$ < 0.2 \%$}
\pdfoutput=1
\newif\ifcolorfigs
\colorfigstrue        

\makeatletter
\def\ps@pprintTitle{%
 \let\@oddhead\@empty
 \let\@evenhead\@empty
 \def\@oddfoot{}%
 \let\@evenfoot\@oddfoot}
\makeatother

\journal{}

\begin{document}

\begin{frontmatter}
\title{Model Independent Approach to the \\Single Photoelectron Calibration of Photomultiplier Tubes}

\author[kavli,efi,chicago]{R. Saldanha\corref{correspondingauthor}\fnref{now}}
\cortext[correspondingauthor]{Corresponding author}
\ead{richard.saldanha@pnnl.gov}
\fntext[now]{Present Address: Pacific Northwest National Laboratory, Richland, WA 99352, USA}
\author[kavli,efi,chicago]{L. Grandi}
\author[fermilab,kavli]{Y. Guardincerri\fnref{dec}}
\fntext[dec]{Deceased} 
\author[kavli,chicago]{T. Wester}
\address[kavli]{Kavli Institute for Cosmological Physics, University of Chicago, Chicago, IL 60637, USA}
\address[efi]{Enrico Fermi Institute, University of Chicago, Chicago, IL 60637, USA}
\address[chicago]{Department of Physics, University of Chicago, Chicago, IL 60637, USA}
\address[fermilab]{Fermi National Accelerator Laboratory, Batavia, IL 60510, USA}

\begin{abstract}
The accurate calibration of photomultiplier tubes is critical in a wide variety of applications for which it is necessary to know the absolute number of detected photons or precisely determine the resolution of the signal. Conventional calibration methods rely on fitting the photomultiplier response to a low intensity light source with analytical approximations to the single photoelectron distribution. We show that this approach often leads to biased estimates due to an inability to model the full distribution accurately, especially at low charge values. We present a simple statistical method to extract the relevant single photoelectron calibration parameters (first two central moments) without making any assumptions about the underlying single photoelectron distribution. We illustrate the use of this method through the calibration of a Hamamatsu R11410 photomultiplier tube and study the accuracy and precision of the method using Monte Carlo simulations. The method is found to have significantly reduced bias compared to conventional methods and works under a wide range of light intensities, making it suitable for the simultaneous calibration of large arrays of photomultiplier tubes where uniform illumination may not be possible.\\

\noindent {\it Dedication:} This paper is dedicated to the memory of Yann Guardincerri, a wonderful colleague and a dear friend. 
\end{abstract}

\end{frontmatter}

\section{Introduction}
Photomultiplier tubes (\pmts) are widely used to detect low levels of light in scientific experiments, medical instruments, and industrial equipment. \pmt\ operation is typically divided into two regimes - photon counting, where the rate of detected photons is small compared to the timing resolution of the detector such that individual photoelectron pulses do not overlap, and signal integration, for light sources of higher intensities where individual photoelectron signals cannot be distinguished. In the latter case, for applications in which the resolution of the signal plays an important role, such as scintillation spectroscopy or pulse shape discrimination, it is critical to obtain an accurate estimate of the total number of generated photoelectrons, as well as the relative standard deviation of the single photoelectron distribution, since these are often the dominant contributors to the resolution of the signal. The response of a PMT is therefore typically calibrated relative to the mean of the charge distribution corresponding to a single photoelectron (\spe). Knowledge of the \pmt~\spe~response is also necessary in order to combine the output signals from several different \pmts~operating at different gains.

In this work we present a simple statistical method to determine the essential parameters of the \spe~response of a photomultiplier tube without making any assumptions about the shape of the \spe~spectrum. Unlike conventional methods which typically rely on fitting the output charge distribution with a model describing the \spe~spectrum, we use the known statistical properties of the \pmt~response to obtain the calibration parameters directly from the charge distribution, without requiring a fit. The general statistical nature of the method allows it to be applied to any kind of photomultiplier tube and to a wide range of illumination levels while only requiring a pulsed light source, a device that is already commonly implemented in scintillation detectors.

\section{Conventional Methods}
The standard method to calibrate the \spe~response of a \pmt~is to use a low intensity light source such that the probability of generating more than a single photoelectron within the time resolution of the detector is small. The resulting spectrum of the integrated signal is then fit with a parameterized model of the \spe~response, in order to obtain the mean and variance for each individual \pmt. The difficulty of such a method lies in the choice of the model. Electron multiplication within the dynode chain is a branching process where the output charge at the \pmt~anode depends on the secondary electron emission probability at each dynode. For the typical photoelectron, i.e. one generated at the photocathode, the most commonly used approximation is a standard Gaussian distribution \cite{Bellamy1994468}, where the mean of the single photoelectron distribution is simply taken as the peak, though more complicated models \cite{Prescott1966173, Lombard1961} are also used to try and accurately model the electron cascade process. Additionally, a large variety of sub-optimal trajectories of photons and electrons through the \pmt~are also possible. For example, a photon may pass through the cathode and directly strike the first dynode \cite{Wright2010, Kaether2012}, a photoelectron may inelastically backscatter off the first dynode \cite{Flyckt2002, Kaether2012} or skip a dynode stage \cite{Burle1980}.  Such trajectories often lead to under-amplified photoelectron signals, increasing the component of the \spe~spectrum with less charge than that at the peak. Since these under-amplified photoelectrons are generated during normal operation and contribute to the total integrated signal, they should be included when estimating the mean and variance of the \spe~response. Under-amplified photoelectrons can account for as much as 20\% of the \spe~spectrum in some models of \pmts, decreasing the mean of the \spe~response by 10-20\% relative to the peak \cite{Haas201106, Dossi2000623}. Ignoring the contribution of under-amplified photoelectrons can lead to an underestimate of the number of detected photoelectrons and an incorrect estimate of the resolution. 

The true shape of the under-amplified component is often difficult to determine due to the large overlap with contributions from electronics noise. Several authors have proposed adding additional terms to the fit function of the single photoelectron response, including a falling exponential, and additional Gaussian components \cite{Dossi2000623, ChirikovZorin2001310}. However, the relative weight and shape of the under-amplified component can vary with the type of photocathode and dynode structure, and can even differ for individual \pmts~of the same model and gain \cite{Wright2010, Haas201106}. Thus it is often difficult to construct a parameterization of the single photoelectron spectrum that is suitable for a range of \pmts~and conditions.
\section{Model-Independent Method}
\begin{figure}[ht!]
\begin{center}
\ifcolorfigs
\includegraphics[width=\columnwidth]{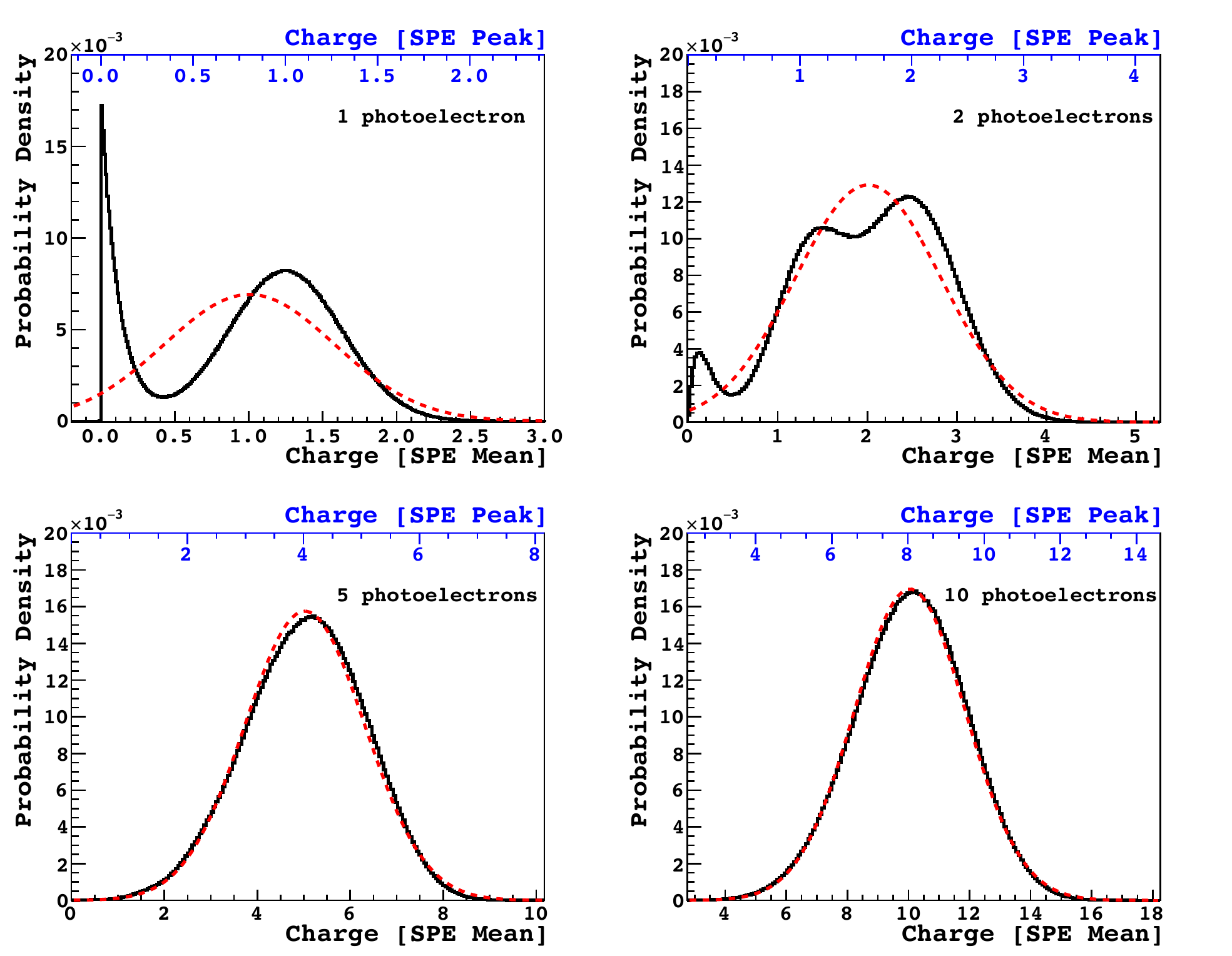}
\else
\includegraphics[width=\columnwidth]{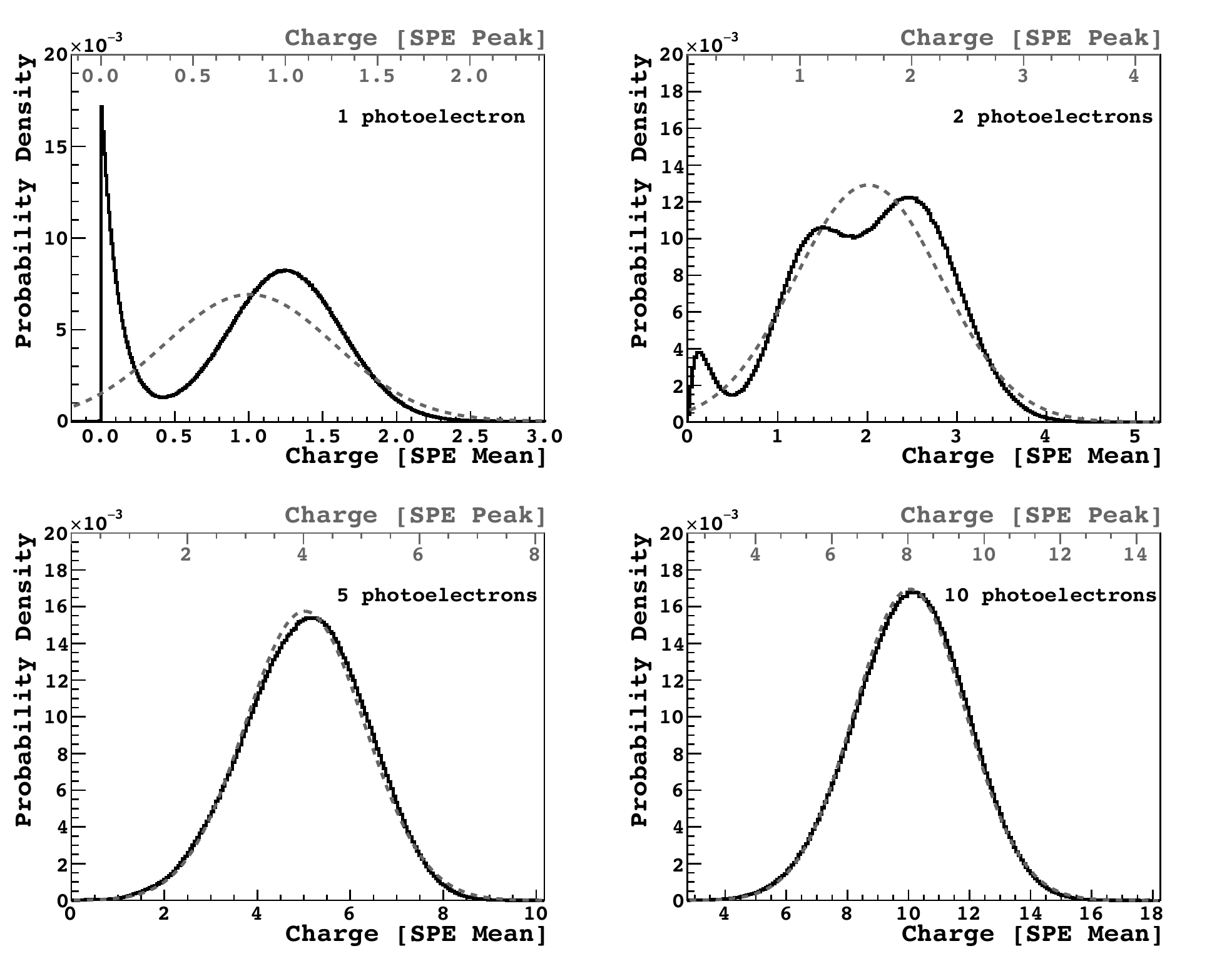}
\fi
\caption{Solid Line: Simulated charge distribution of 1, 2, 5 and 10 photoelectrons for a \pmt~with a Gaussian distribution of fully-amplified photoelectrons and a large exponential component of under-amplified photoelectrons. Background noise was not included in the simulation. Dashed Line: Gaussian distribution with the same total mean and variance as the corresponding \pmt~charge distribution. The bottom axis shows the charge calibrated with respect to the true \spe~mean, while the top axis is calibrated with respect to the peak of the fully-amplified photoelectrons.}
\label{fig:clt_example}
\end{center}
\end{figure}
The single photoelectron calibration method presented here focusses on accurately estimating the mean and variance of the \spe~distribution at the PMT output, including contributions from underamplified photoelectrons, without making any assumptions about the shape of the distribution. For most experimental purposes, knowledge of the higher moments, or the entire functional form of the \spe~response is not required. We assume that the \pmt~will be used within its linear regime and this ensures that, as the number of photoelectrons increases, the response to multiple photoelecrons quickly converges (by the central limit theorem (\clt)) to a Gaussian distribution that is completely described by the first two central moments of the single photoelectron response. This is illustrated in Figure~\ref{fig:clt_example}, which shows the charge spectra from 1, 2, 5, and 10 photoelectrons where each photoelectron is randomly and independently drawn from an \spe~distribution with a large fraction of under-amplified photoelectrons. Even in this case the response to $n \geq 5$ photoelectrons (\si{\pe}) is to a very good approximation Gaussian, with a mean and variance $n$ times that of the entire \spe~distribution. Though the distribution converges to a Gaussian for large number of photoelectrons, it is critical to include the under-amplified component in the estimate of the mean and variance in order to obtain the correct estimate of the photoelectron statistics in the signal. For example, if one were to calibrate the \spe~mean as simply the Gaussian peak position of the \spe~distribution in the top left panel of Figure~\ref{fig:clt_example}, one would obtain an SPE mean a factor 1.25 higher than the true mean. Using this incorrect value would result in a 20\% [(1.25-1)/1.25] underestimate of the true number of photoelectrons in the other panels (or any given signal). We note that the above considerations also apply when the light is distributed over an array of photomultipliers, and the total signal is obtained by summing the output of all the PMTs. In this case, even though the \spe~responses of the individual \pmts~are not necessarily identical, variants of the classical \clt~typically ensure convergence to a Gaussian distribution. Thus, for scintillation signals that produce more than 5 \si{\pe} on average, the contribution of the photomultiplier response is determined by only the first two central moments of the \spe~distribution.\footnote{For smaller signals, the knowledge of the mean and variance are still necessary for calibration, and are still accurately estimated by the method described in this paper, though higher moments may also need to be calculated to fully model the shape of the detector response.} 

In order for the description of the method to be clear, we must first briefly describe the corresponding experimental setup. A low intensity, pulsed laser is used to illuminate the \pmt~to be calibrated. The laser is externally triggered and for each trigger the PMT output is integrated at the time corresponding to the expected anode output signal. We stress that in this setup the \pmt's~output for every trigger is recorded, even if there is no visible signal. This ensures that no under-amplified photoelectrons are missed.

For every trigger, there are two contributions to the total measured charge $q$, one from the background noise that is always present in the system, regardless of the presence or absence of a laser-induced photoelectron signal, and one associated with the presence of a photoelectron signal. The total integrated charge is simply the sum of these two contributions, which, by definition, are independent. We will denote the probability distribution of the total integrated charge as \totaldist, and the background and signal probability distributions as \bkgddist~and \signaldist~respectively. It then follows that
\begin{linenomath}
\begin{align}
\totaldist = (\bkgdrand \conv \signalrand)(q)
\end{align}
\end{linenomath}
where $\conv$ indicates a convolution of the two distributions.
For independent random variables, the first two moments are additive, and hence
\begin{align}
\label{eq:total_mean}
\mean{\totalrand} = \mean{\bkgdrand} + \mean{\signalrand}\\
\label{eq:total_var}
\var{\totalrand} = \var{\bkgdrand} + \var{\signalrand}
\end{align}
where \mean{X} and \var{X} denote the mean and variance of the distribution $X$ respectively, and we have omitted the domain of the distributions for clarity.

The signal charge distribution can be written in terms of the number of photoelectrons $p$ produced
\begin{linenomath}
\begin{align}
\signaldist = \sum_{p=0}^\infty \signaldistv{p}\pedist
\end{align}
\end{linenomath}
where $\signaldistv{p}$ is the signal charge distribution corresponding to exactly $p$ photoelectrons and \pedist~is the discrete probability distribution of the number of photoelectrons produced in a single laser pulse. We shall denote the charge distribution of the \spe~response as $\signaldistv{1} \equiv \spedist$. Assuming that the PMT response is linear, the multi-photoelectron response \signaldistv{p} is the p-times repeated convolution of \spedist, denoted as $\signaldistv{p} \equiv \spedistconv{p}$. The mean and variance of these two distributions are related by \mean{\sperandconv{p}} = $p\cdot \mean{\sperand}$ and \var{\sperandconv{p}} = $p\cdot \var{\sperand}$. Using these two properties one can calculate the mean and variance of the signal
\begin{linenomath}
\begin{align}
\mean{\signalrand} &= \mean{\sperand} \cdot \mean{\perand}\\
\var{\signalrand} &= \var{\sperand} \cdot \mean{\perand} + \meansq{\sperand} \cdot \var{\perand}
\end{align}
\end{linenomath}
Finally, we can substitute the above in \eqns\eqref{eq:total_mean} and \eqref{eq:total_var} to obtain the first two central moments of the single photoelectron response $\psi(q)$
\begin{linenomath}
\begin{align}
\label{eq:spe_mean_final}
\mean{\sperand} &= \frac{\mean{\totalrand} - \mean{\bkgdrand}}{\mean{\perand}} \\
\label{eq:spe_var_temp}
\var{\sperand} &= \frac{\var{\totalrand} - \var{\bkgdrand} - \meansq{\sperand}\cdot \var{\perand}} {\mean{\perand}}
\end{align}
\end{linenomath}
As can be seen from the above equations, in order to obtain the mean and variance of the \spe~distribution, one needs to know the mean and variance of the photoelectron distribution $L(p)$. 

For an ideal laser emitting coherent light in a single mode, the distribution of the number of photons follows a Poisson distribution \cite{Saleh1991}, with the variance equal to the mean. Even in the case where the emitted light is not perfectly Poissonian, it can be shown that after a random deletion process (such as attenuation by optical filters or conversion to photoelectrons with non-unity quantum efficiency) the output distribution approaches a Poisson distribution with reduced mean  \cite{Hu2007173, Teich82}. Explicitly, if the initial photon distribution has a ratio of the variance to the mean (defined as the Fano factor), $F_i$, then after attenuation by a factor $\eta$, the Fano factor $F_o$ of the output distribution is
\begin{linenomath}
\begin{align}
F_o = 1 + \frac{(F_i -1)}{\eta} \notag
\end{align}
\end{linenomath}
Thus even for non-ideal light sources, if the output is strongly attenuated ($\eta \gg 1$), $F_o \approxim 1$ and the variance approaches the mean, as for a Poisson distribution. Given these considerations, we can assume that the distribution of detected photons from a strongly attenuated laser light source follows a Poisson distribution. We can therefore further simplify \eqn\eqref{eq:spe_var_temp} by setting the variance of the photoelectron distribution equal to the mean, \var{\perand} = \mean{\perand}, to get:
\begin{linenomath}
\begin{align}
\label{eq:spe_var_final}
\var{\sperand} &= \frac{\var{\totalrand} - \var{\bkgdrand}} {\mean{\perand}} - \meansq{\sperand}
\end{align}
\end{linenomath}
Before discussing the method to estimate the parameters on the right hand side of \eqns \eqref{eq:spe_mean_final} and~\eqref{eq:spe_var_final} it is worthwhile to explicitly list some of the assumptions made above and compare them to other methods of single photoelectron calibration.
\begin{itemize}
\item Unlike fitting methods, we have not assumed any functional form for the \spe~response. The above equations are valid for any \spe~distribution with no assumptions about the shape or amplitude of the under-amplified photoelectron distribution. 
\item Similarly, we have also made no assumption about the shape of the background noise distribution, which is determined by the specific electronics used and the background noise present in the setup.
\item As with most fitting methods, we have assumed that the PMT and any associated electronics respond linearly to the number of photoelectrons. In the typical regime used by this method, fewer than 20 \si{\pe} are produced in each laser pulse, which is well within the linear range of most PMTs.
\item The above formulation divides the contributions of the total charge into two categories, background and signal. The background distribution accounts for all signals that are independent of the photoelectron production by the laser. This includes any noise from the electronics, the trigger and the pulsing of the laser (which occurs for every trigger) as well as dark count signals produced by thermionic emission from the photocathode and dynode chain. The signal distribution is assumed to only include contributions that are linear with the number of laser-induced photoelectrons. In certain experimental setups, there may be contributions that do not fall into either category. For example, noise from a discriminator firing (when the signal is above a certain threshold) may only occur when the laser light produces a signal, but it does not increase as the number of photoelectrons increases. In such cases, as with other calibration methods, care must be taken to account for these contributions in the total charge distribution. 
\item The shape and relative contribution of under-amplified photoelectrons to the \spe~spectrum can depend on various factors such as external magnetic fields, non-uniformity of the photocathode, the intensity and angular distribution of light incident on the photocathode, etc. As with all calibration methods, it is therefore important that the setup and illumination of the PMT during calibration be as close as possible to that of the PMT during its regular operation.
\end{itemize}  
\section{Experimental Setup}
\begin{figure}[tbp]
\begin{center}
\includegraphics[width=\columnwidth]{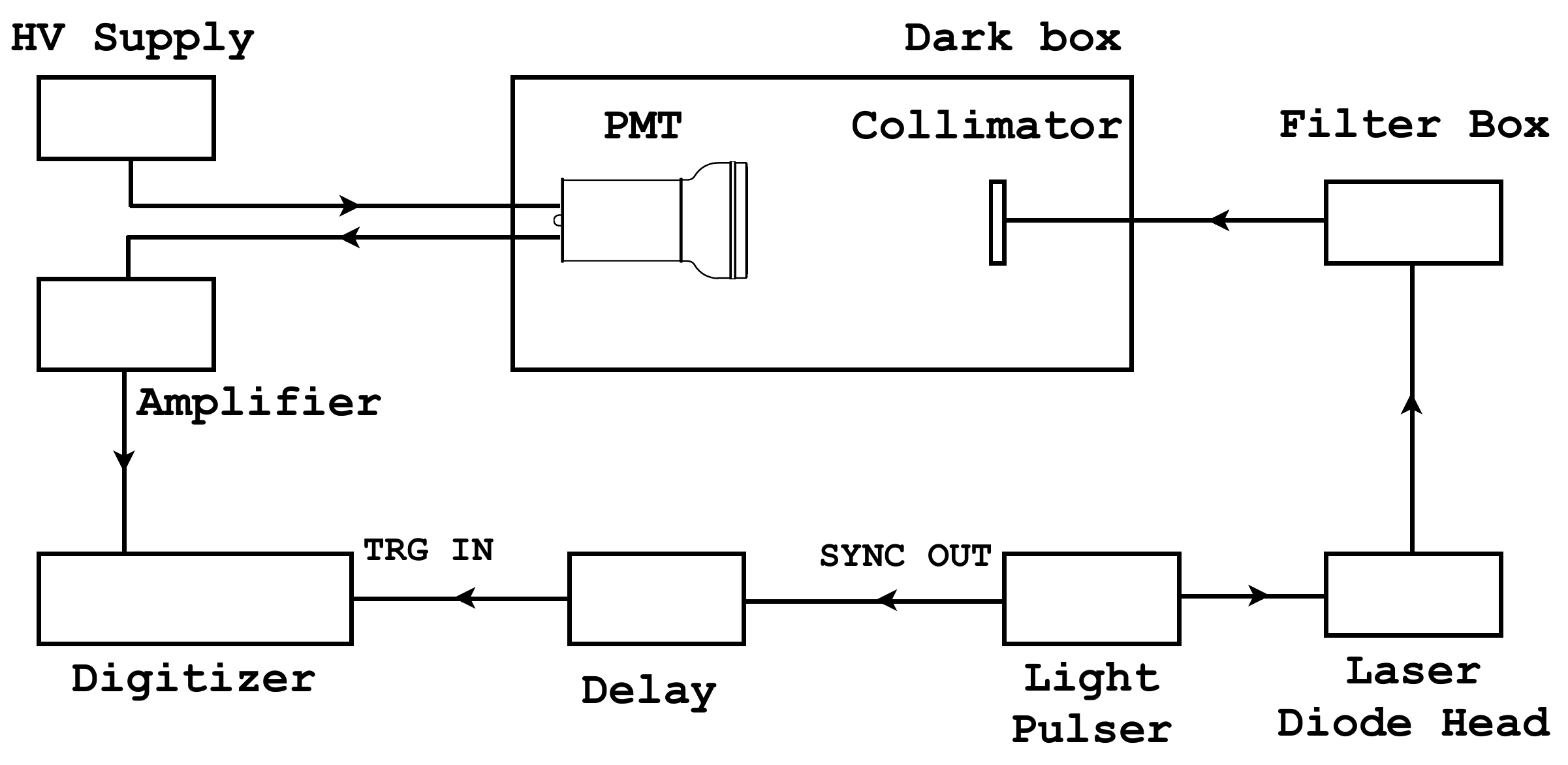}
\caption{Schematic diagram of the experimental setup used to measure the single photoelectron spectrum with a pulsed laser light source.}
\label{fig:exp_setup}
\end{center}
\end{figure}

In order to test the calibration method described above we have used the experimental setup illustrated in Figure~\ref{fig:exp_setup}. 
A Hamamatsu \pmtmodel~3" photomultiplier with a 12 stage box and linear-focused dynode structure \cite{Hamamatsu} was placed in a custom-made 
metal darkbox that featured a continuous conducting surface to reduce the effect of electrical noise. The~cathode was maintained at negative high voltage 
with a total divider resistance of \dividerresistance~between the cathode and the anode, and the recommended voltage distribution ratio\footnote{4:1.5:2:1:1:1:1:1:1:1:1:2:1} 
between the dynode stages. The \pmt~was illuminated by a collimated optical fibre that carried light from a fast pulsed laser diode (Hamamatsu PLP-10-040C \cite{Hamamatsu}) which we will henceforth refer to as a laser. 
The laser pulses had a typical width of \laserpulsewidth~(FWHM) and a wavelength of \laserwavelength. 
The intensity of the light incident on the \pmt~was varied by placing different neutral density filters along the light path. 
For all data sets acquired, the combined attenuation factor $\eta$ of the filters was kept $\geq 10^{5}$, in order to ensure 
that the photon distribution was Poissonian. 
The anode of the PMT was terminated with a \terminatorresistance~resistor and connected to a custom fast amplifier with a \ampgain~gain. 
The output of the amplifier was then sent to a \digitizerbit, \digitizersamplerate~CAEN \digitizermodel~digitizer \cite{Caen}. 
The digitizer was externally triggered by the synchronous output of the laser, delayed by \triggerdelay~with respect to the optical signal; this avoids any noise related to the triggering of the digitizer from overlapping with the PMT output in the time window of interest. For each trigger a \laseracquiwindow~digitized 
waveform was recorded and stored for analysis offline. 

\begin{figure}[t]
\begin{center}
\ifcolorfigs
\includegraphics[width=\columnwidth]{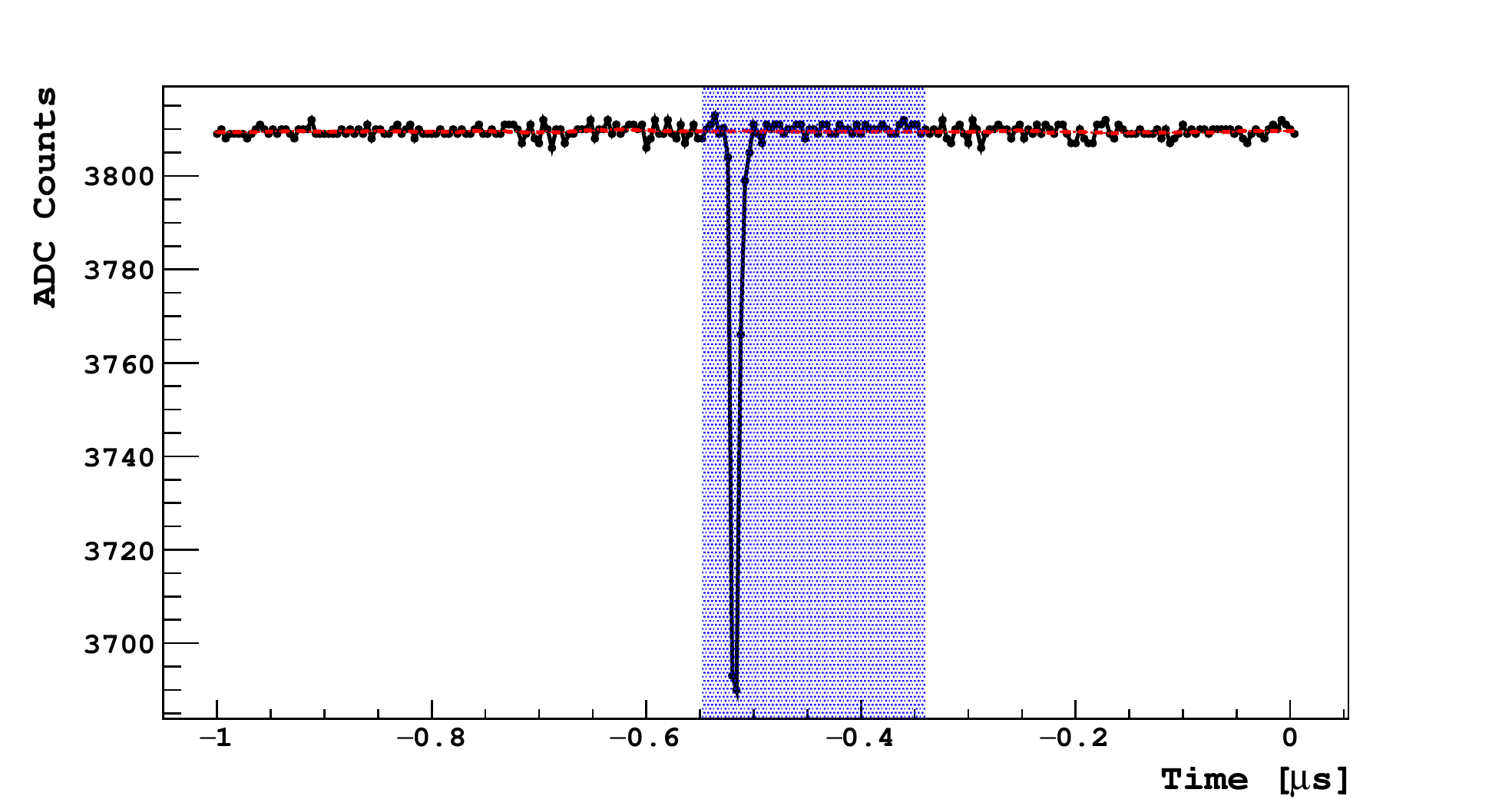}
\else
\includegraphics[width=\columnwidth]{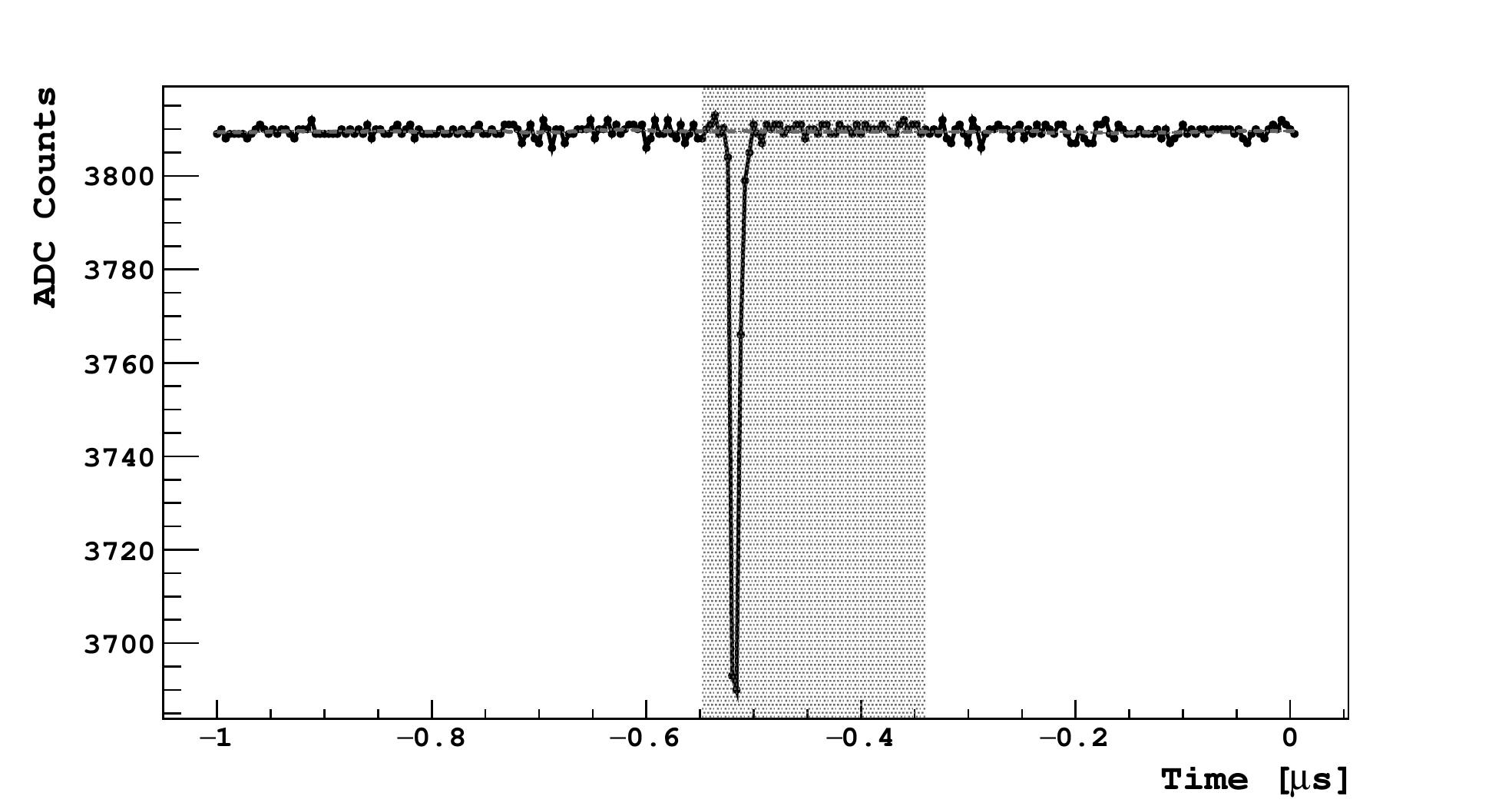}
\fi
\caption{Example of a digitized PMT signal within a sample trigger window, acquired during a \laserdata~data set. 
The black points indicate the digitized waveform, with the dashed horizontal line showing the estimated baseline. 
The shaded region indicates the fixed laser time window chosen for integration. 
The negative pulse in the laser time window likely indicates the presence of a laser-induced photoelectron signal.}
\label{fig:samp_waveform}
\end{center}
\end{figure}
For each configuration of light intensity and \pmt~voltage that was studied, 
two data sets of $\nsamps = 500,000$ triggers were acquired at a trigger rate of \laserpulserate. 
A ``\laserdata"~data set was acquired with the optical fibre connected such that the laser light illuminated 
the \pmt~and another ``\blankdata"~data set was acquired with the optical fibre disconnected before the filter box, 
and the fibre feedthrough capped. As will be described in detail in the next section, 
the \laserdata~data set will be used to estimate the moments of the total charge distribution (signal + background), 
the \blankdata~data set will be used to estimate the moments of the background charge distribution,
and the combination of both will be used to estimate the photoelectron distribution. 
In order to ensure that the noise levels remained the same for both the \laserdata~and \blankdata~data, 
all the electronics (including the laser) and wiring were kept in the same operating conditions for both runs. 
As a consistency check, for a few configurations, a \blankdata~data set was taken before and after the \laserdata~data set 
and the integrated charge distribution for the two \blankdata~data sets were compared using the Kolmogorov-Smirnov test. 
In all cases, the integrated charge spectrum of the \blankdata~data sets were found to be compatible (p-value $> 0.1$).

The event reconstruction required for this calibration method is straightforward. 
The expected time window for the laser-induced PMT signal is identified empirically by averaging together all the waveforms acquired during a \laserdata~data set and then selecting a \laserintegrationwindow~time window to include the entire laser-induced signal. Note that the chosen window should start early enough to include signals from photons passing through the photocathode and directly striking the first dynode \cite{Kaether2012}. 
Outside of the laser time window a baseline is calculated for each individual waveform using a moving average of $\pm$ \baselinewindow~around 
each sample. The baseline within the laser time window is then linearly interpolated using the samples on either side of the window. 
This method ensures that the baseline is evaluated in the same way regardless of whether or not a laser signal is present. 
A sample waveform along with the defined laser time window and estimated baseline is shown in Figure~\ref{fig:samp_waveform}.  

\begin{figure}[t!]
\centering
\ifcolorfigs
\includegraphics[width=\columnwidth]{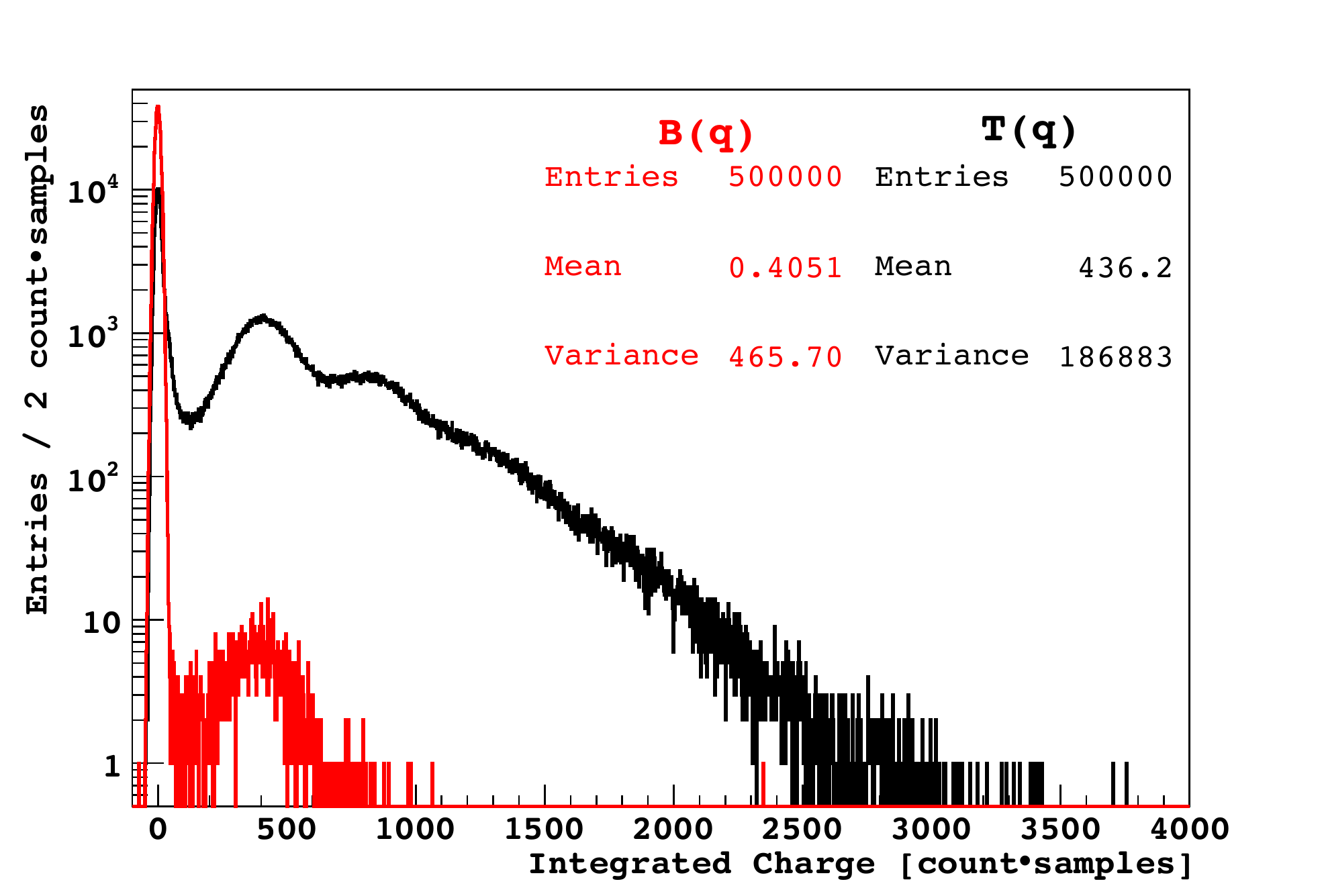}
\else
\includegraphics[width=\columnwidth]{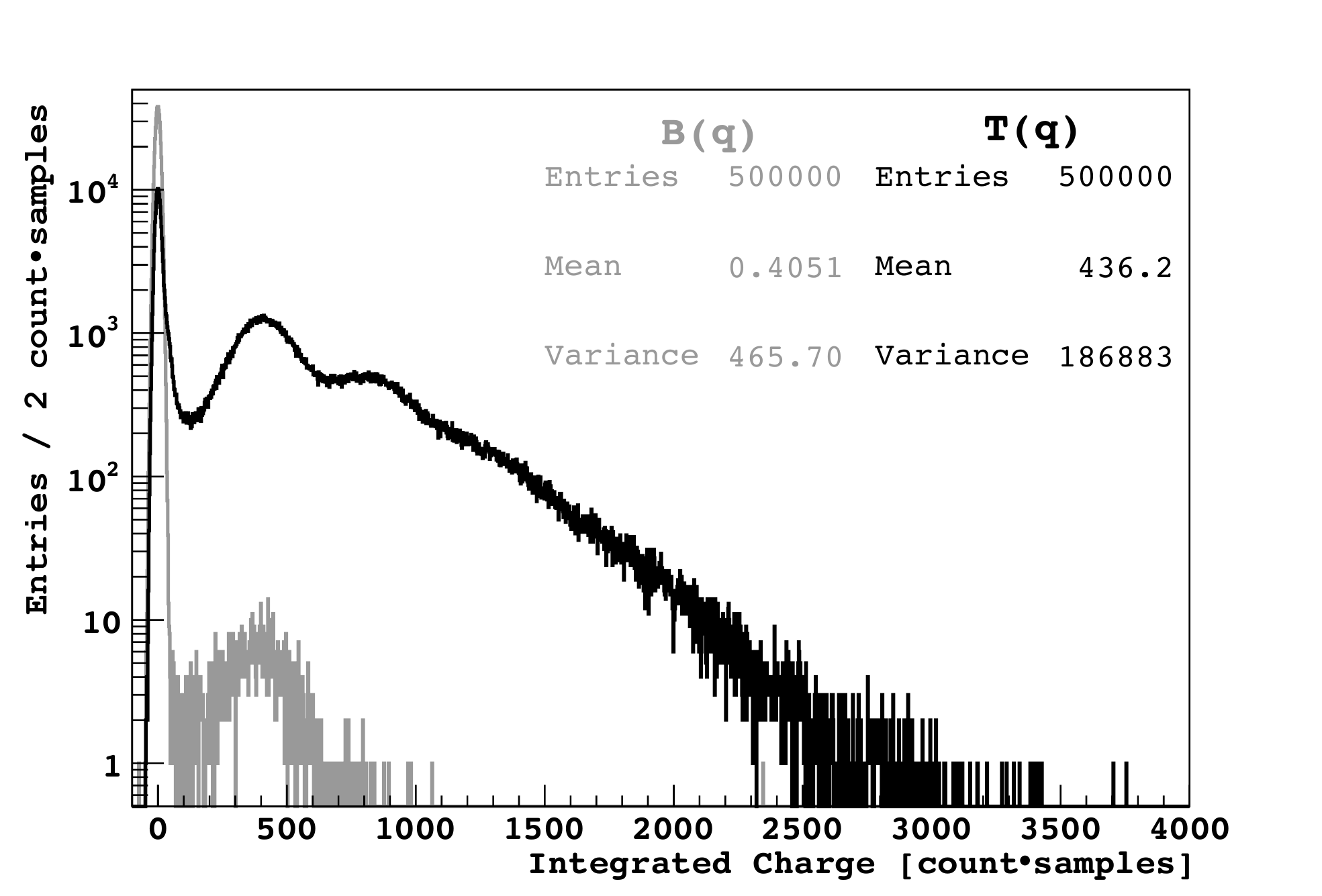}
\fi
\caption{Integrated charge spectra acquired at an absolute PMT voltage difference of 1700 V. The \laserdata~data spectrum, \totaldist, had an estimated occupancy of 1.37 photoelectrons/trigger, while the \blankdata~data spectrum, \bkgddist, was acquired at the same settings as the \laserdata~data set, 
but with the optical fibre disconnected.}
\label{fig:integral_hist}
\end{figure}

The integral (inverted to account for the negative PMT pulses) of the baseline-subtracted waveform over the defined laser time window is 
calculated for each trigger. Figure~\ref{fig:integral_hist} shows the distribution of the integral for a \laserdata~and a \blankdata~data set, 
acquired at an absolute PMT voltage difference of 1700 V (\approxim~\pmtgainseventeen~gain) and a filter attenuation of $5\times10^{-6}$. 
In both distributions the peak centered at zero is primarily due to fluctuations of the noise about the estimated baseline 
with no photoelectron signal present. The peak at \SI{400}{\countsamples} in the \laserdata~data is due to fully amplified single photoelectrons from the photocathode, and a peak due to two fully amplified photoelectrons at \SI{800}{\countsamples} is also visible. 
The small peak at \SI{400}{\countsamples} in the \blankdata~data is due to dark noise photoelectrons and possibly small amounts of stray light entering the darkbox (\darknoisefraction~of triggers) that accidently fall within the laser time window. 
Since the spectrum of dark-noise photoelectrons does not necessarily follow the same distribution as photon-induced photoelectrons \cite{Dossi2000623, Flyckt2002}, the presence of these events in both the \blankdata~and \laserdata~data allows us to correctly account for them and exclude them from the estimation of the single photoelectron mean.
\section{Parameter Estimation}
\label{sec:par_estimation}


As can be seen from \eqns\eqref{eq:spe_mean_final} and~\eqref{eq:spe_var_final}, 
in order to determine the first two central moments of the SPE response we need to evaluate 
the first two central moments of the total charge distribution and the background distribution, 
as well as the mean number of photoelectrons produced in each trigger. 
Since we do not have prior knowledge of the true underlying distributions, we will estimate the moments 
from the experimentally measured data sample of \nsamps~triggers.

The central moments of the total charge distribution can be directly obtained by calculating the mean and variance of the measured PMT output spectrum in the presence of the laser, which is typically the spectrum that is used to fit the \spe~response in other methods. 
An example spectrum is shown in Figure~\ref{fig:integral_hist}, along with the calculated mean \mean{\totalrand}, 
and variance \var{\totalrand}, to be used in \eqns \eqref{eq:spe_mean_final} and~\eqref{eq:spe_var_final}.  

There is often an overlap of the background distribution and the signal distribution of under-amplified single photoelectrons. 
This makes it difficult to cleanly determine the mean and variance of the background in the presence of the laser signal. 
For this reason we estimate the moments of the background spectrum from the separately measured \blankdata~data set, 
an example of which is shown in Figure~\ref{fig:integral_hist}, along with the calculated mean \mean{\bkgdrand}, and variance \var{\bkgdrand}. 

The only parameter that is not straightforward to estimate is the mean number of laser-induced photoelectrons produced in each trigger, 
$\mean{\perand}$, which we shall refer to as occupancy. 
As discussed previously, the number of photoelectrons produced follows a Poisson distribution, which can be written as
\begin{linenomath}
\begin{align}
\pedist &= \frac{\occup^p e^{-\occup}}{p!}\\
\occup &\defined \mean{\perand} = \var{\perand} 
\end{align}
\end{linenomath}
The occupancy $\lambda$ is directly related to the probability of producing zero laser-induced photoelectrons, 
\begin{linenomath}
\begin{align}
\occup &= -\ln{\left(\pedistzero\right)}
\end{align}
\end{linenomath}
which can be estimated from the number of sample triggers with zero laser-induced photoelectrons (\zerope~triggers), \nsampszero, and the total number of sample triggers \nsamps
\begin{linenomath}
\begin{align}
\estim{\occup} &\equiv -\ln{(\estim{\nsampszero}/\nsamps)}
\end{align}
\end{linenomath}
where \estim{\occup}~and \estim{\nsampszero}~denote the estimates of the occupancy and number of \zerope~triggers in the \laserdata~data sample respectively.

There are several different techniques that can be used to estimate the value \nsampszero~and 
the optimal method will depend on the nature of the signal and background distributions. 
For example, if the temporal shape of the PMT output pulse is known, and the triggers are individually recorded, 
one can assign a likelihood for the presence of a laser-induced signal to each individual trigger. 
For the purposes of this paper, we will restrict ourselves to a very simple algorithm, whose statistical and systematic uncertainties can be estimated analytically.

We will use the fact that we have access to a pure sample of \zerope~events from the \blankdata~data set and hence have empirical information about the shape of the \zerope~distribution.\footnote{Note that while some \blankdata~data events may contain photoelectron signals due to stray ambient light or dark noise, by construction all events have zero laser-induced photoelectrons.} Triggers in the \laserdata~data set that contain a non-zero number of laser-induced photoelectrons typically have a higher charge output than triggers that only contain background noise. We can therefore use the comparison of the number of events in low-charge region of the \laserdata~and \blankdata~spectrum to estimate the number of \zerope~triggers as follows: 
\begin{enumerate}
\item We place a threshold cut at a low charge value such that the fraction of \laserdata~triggers with a non-zero number of laser-induced photoelectrons that fall below the cut is expected to be small (we will quantify this requirement in \sect\ref{subsec:sys_unc}). We count the number of triggers, \nlaseramp, in the \laserdata~data set below the threshold and assume they are only zero-pe triggers. 
\item In order to get the total number of \zerope~triggers in the \laserdata~sample, \nsampszero, we use the shape of the \blankdata~spectrum to correct the value of \nlaseramp~for the number of \zerope~triggers that fall above the threshold. We define the fraction of \blankdata~data samples that fall below the threshold cut as \zeropefrac. The estimated total number of \zerope~triggers in the \laserdata~data set is then
\begin{linenomath}
\begin{align}
\estim{\nsampszero} &= \frac{\nlaseramp}{\zeropefrac}
\end{align}
\end{linenomath}
\end{enumerate}
Our estimate for $\estim{\occup}$ is therefore 
\begin{linenomath}
\begin{align}
\estim{\occup} &= -\ln{(\nlaseramp/\zeropefrac\nsamps)}
\end{align}
\end{linenomath}
Rather than expressing the position of the threshold cut in terms of a charge value, which will depend on the shape of the background distribution, we will express it in terms of \zeropefrac, the fraction of the background distribution that falls below the threshold. This not only lets us describe the choice of threshold in a way that can be easily translated to different experimental setups, but, as we will see in the next sections, the statistical and systematic uncertainties are also conveniently expressed in terms of the fraction, \zeropefrac.
\section{Parameter Uncertainties}
\label{sec:par_unc}
\begin{figure}[t!]
\centering
\includegraphics[width=\columnwidth]{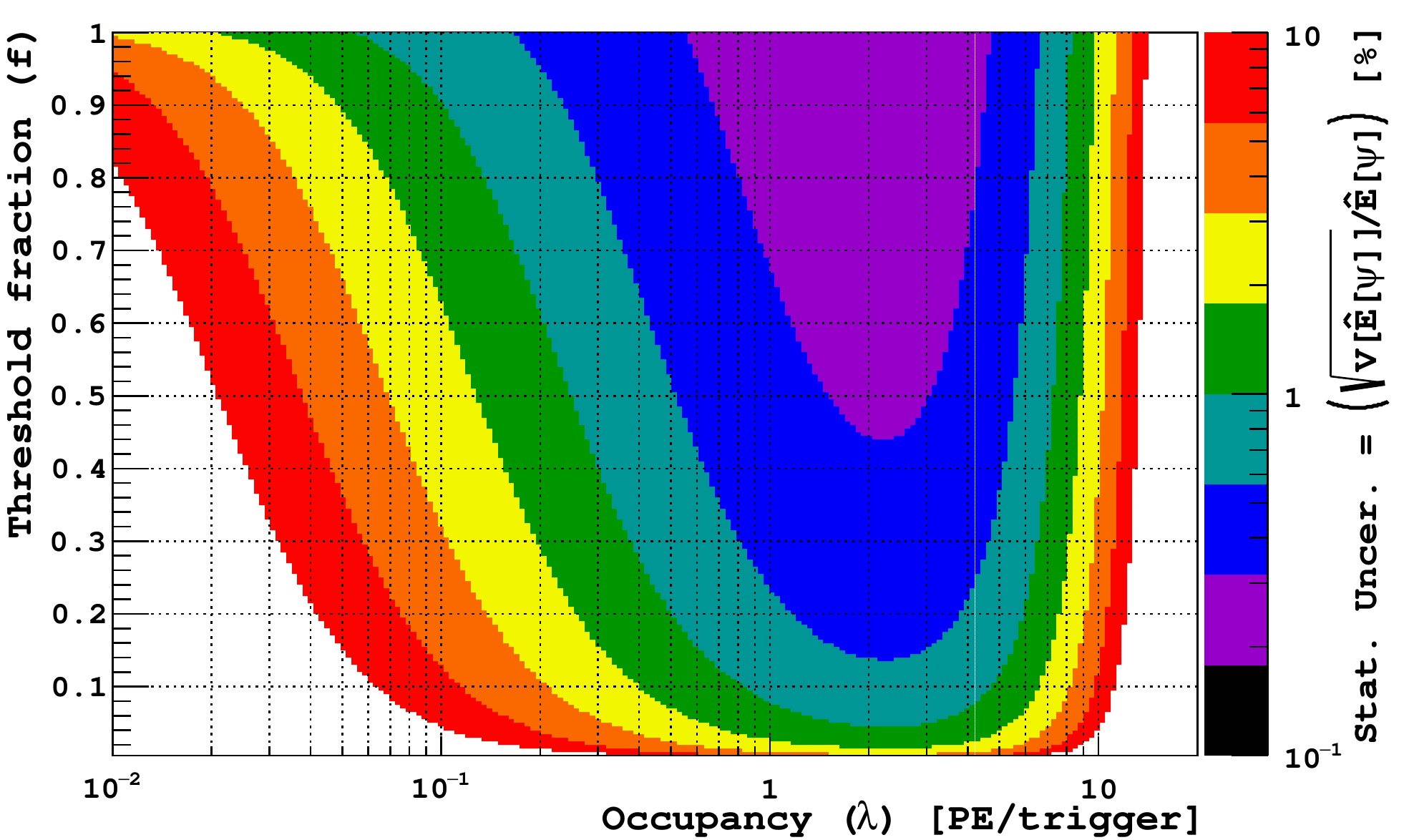}
\caption{Statistical uncertainty of the model independent method as a function of the occupancy and the fraction of the background spectrum below the threshold cut. Values are calculated using \eqn\eqref{eq:stat_unc_final} with the parameters of our experimental setup.}
\label{fig:precision_occ_f}
\end{figure}
In order to find the optimal operating parameters for the laser intensity and number of triggers, 
it is useful to calculate the statistical and systematic uncertainties corresponding to the estimate of the moments from the finite data sample.

\subsection{Statistical Uncertainties}
We can calculate the statistical fluctuations in the estimate of the single photoelectron mean as (see \append\ref{sec:stat_unc_occupancy} for derivation)
\begin{linenomath}
\begin{align}
\label{eq:stat_unc_final}
\var{\sampmean{\sperand}} \approxim&~\frac{\occup(\meansq{\sperand}+\var{\sperand}) + 2\var{\bkgdrand}}{\nsamps \occup^2} \notag\\
&+ \frac{\meansq{\sperand}\left(e^{\occup}+ 1 - 2\zeropefrac\right)}{\zeropefrac \nsamps \occup^2}
\end{align}
\end{linenomath}
where \sampmean{\sperand} denotes the estimated \spe~mean, $i.e.$ the estimate of the mean of the \spe~distribution from the finite data sample acquired. 
The statistical uncertainty of the estimated single photoelectron mean decreases as the number of trigger samples \nsamps~increases, with the optimal value for the occupancy depending on the given photomultiplier and 
the background spectrum. Figure~\ref{fig:precision_occ_f} shows how the statistical uncertainty of the  \spe~mean estimate, 
evaluated with \eqn\eqref{eq:stat_unc_final}, varies as a function of the occupancy and \zeropefrac~for the parameters of our experimental setup.
The statistical uncertainty has a broad minimum at an occupancy value of \occup\approxim2 and decreases as  \zeropefrac, the fraction~of background events falling below the chosen threshold, increases. 
The second term in \eqn\eqref{eq:stat_unc_final}, associated with the statistical uncertainty in the estimation of the occupancy, dominates the overall statistical uncertainty; hence, the precision of this method is strongly dependent on the precision with which the occupancy can be determined.

The statistical uncertainty in the estimated single photoelectron variance is difficult to express analytically and 
was therefore evaluated using the simulations described in \sect\ref{sec:monte_carlo}. It is largely dominated by the statistical uncertainty in the estimation of the occupancy and can be approximated as
\begin{linenomath}
\begin{align}
\label{eq:var_stat_unc_final}
\var{\sampvar{\sperand}} \approxim&~ \left(\frac{\partial \sampvar{\sperand}}{\partial \sampmean{\perand}} \right)^2 \var{\sampmean{\perand}} + \ldots \notag\\
=&~  \frac{\left(\meansq{\sperand}-\var{\sperand}\right)^2\left(e^{\occup}+ 1 - 2\zeropefrac\right)}{\zeropefrac \nsamps \occup^2}
\end{align}
\end{linenomath}
where we have only considered the dominant term in the first-order Taylor expansion.
\subsection{Systematic Uncertainties}
\label{subsec:sys_unc}
The dominant systematic uncertainty in the estimate of the \spe~mean arises from the evaluation of the occupancy. 
The calculation of the occupancy is made under the assumption that the number of \nonzerope~triggers falling below the \thresholdcut~cut is negligible. However measurements of signals from photomultiplier tubes operated at high gain have shown contributions from under-amplified photoelectrons of arbitrarily small charge \cite{Wright2010, Haas201106}. 
Triggers with a laser-induced photoelectron that produces a very small integrated charge can fall below the \thresholdcut~cut and be incorrectly included in the calculated number of \zerope~triggers \nlaseramp. 
This leads to a systematic decrease in the estimated occupancy \estim{\occup}, and correspondingly a systematic increase 
in the estimated \spe~mean.

The estimated \spe~mean, \biassampmean{\sperand}, averaged over a large number of measurements, \mean{\biassampmean{\sperand}}, can be expressed in terms of the true \spe~mean, \mean{\sperand}, as (see \append\ref{sec:bias})
\begin{linenomath}
\begin{align}
\label{eq:sys_unc_final}
\mean{\biassampmean{\sperand}} &\approxim \mean{\sperand}\cdot \left(1 + \frac{\speleakfrac}{\zeropefrac}\right)
\end{align}
\end{linenomath}
where $\speleakfrac/\zeropefrac$ is the ratio of \speleakfrac, 
the fraction of single photoelectron triggers falling below the \thresholdcut~cut, to \zeropefrac, 
the fraction of the  \blankdata~background spectrum below the cut. 
The value of \speleakfrac~depends not only on the the location of the \thresholdcut~cut, but also on the shape of the \spe~spectrum and the operating gain of the \pmt.
\speleakfrac~decreases as the gain increases and typically decreases faster than $f$ as the position of the \thresholdcut~cut is lowered. Therefore, in order to minimize the systematic bias in the estimate of the \spe~mean, as high a gain as experimentally possible should be used and the value of the \thresholdcut~cut must be chosen such that $\speleakfrac/\zeropefrac$ is small.

It is difficult to calculate the systematic uncertainty in the estimation of the \spe~variance and therefore, like the statistical uncertainty, it is also evaluated using simulations. 
\subsection{Choice of Threshold}
\label{sec:thresh_choice}
\begin{figure}[t!]
\centering
\ifcolorfigs
\includegraphics[width=\columnwidth]{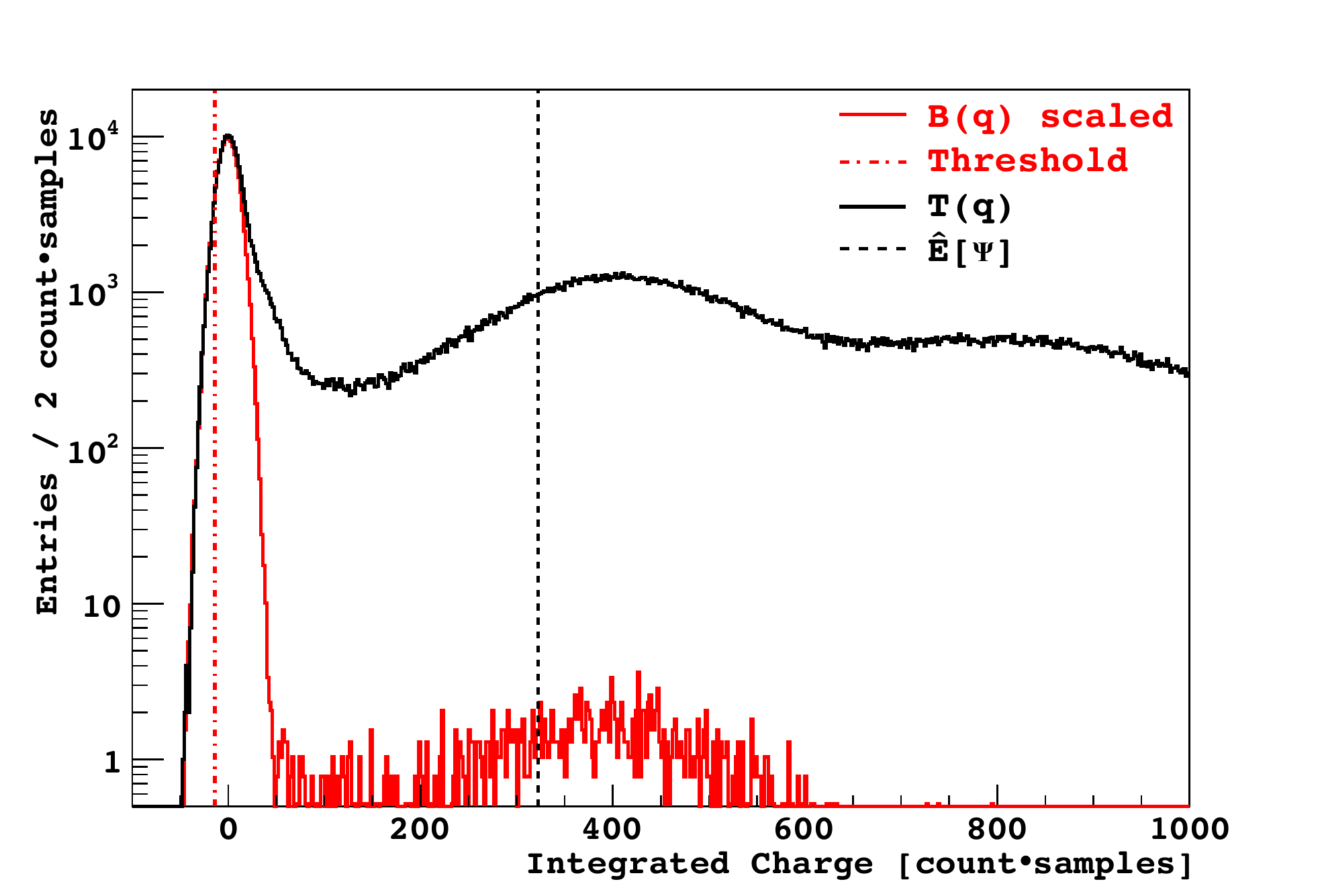}
\else
\includegraphics[width=\columnwidth]{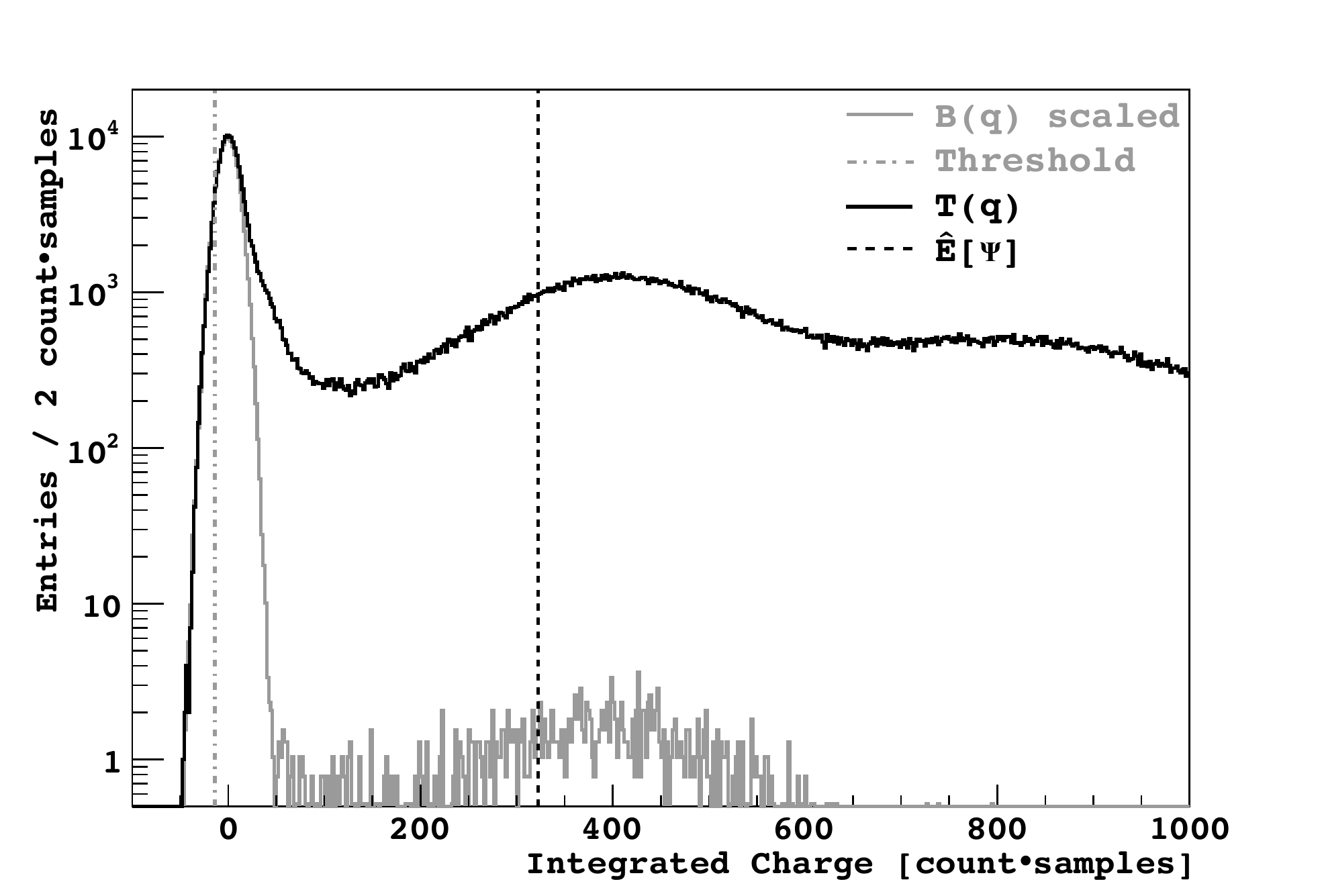}
\fi
\caption{Same data as in Figure~\ref{fig:integral_hist}, with the \blankdata~data spectrum vertically scaled to match the \laserdata~data spectrum 
using the estimated occupancy. The dot-dashed vertical line indicates the position of the threshold cut, corresponding to $\zeropefrac = 0.1$, 
used to estimate the occupancy. The dashed vertical line indicates the value of the estimated single photoelectron mean.}
\label{fig:integral_hist_scaled}
\end{figure}
As discussed above, the position of the \thresholdcut~cut plays a key role in determining both the statistical and 
systematic uncertainty of the results. 
The position should be chosen to keep  \speleakfrac~(the fraction of the single photoelectron triggers falling below the cut) small in order to reduce the systematic bias, 
while still maintaining enough statistics \zeropefrac~(fraction below the cut in the \blankdata~spectrum) to precisely estimate the occupancy. 
If experimentally possible, one should use \eqn\eqref{eq:stat_unc_final} 
(depicted for our specific experimental setup in Figure~\ref{fig:precision_occ_f}) to choose the occupancy 
such that the required precision can be achieved with as low a \thresholdcut~cut as possible, 
since the fractional bias \speleakfrac/\zeropefrac~typically decreases as the cut value is decreased.  
If additional precision is desired, one can always increase the total number, \nsamps, of trigger samples acquired. 

In order for the statistical uncertainty to remain below 3\% for both the data acquired in our experimental setup and 
our Monte Carlo simulations, we have chosen our threshold cut to be $\zeropefrac = \zeropefraclowthreshold$ for occupancies in the range of $0.2 < \occup~[\si{\pepertrigger}] < 8$. For lower or higher occupancies
we have chosen $\zeropefrac = \zeropefrachighthreshold$. Note that we have chosen to always place the cut such that $\zeropefrac < 0.5$, corresponding to an integrated charge below zero (see for example the dot-dashed line in Figure~\ref{fig:integral_hist_scaled}). This implies that only very small single photoelectron signals 
(which when summed with the background noise lead to an overall negative charge) will fall below the \thresholdcut~cut.
\section{Experimental Results}
\label{sec:exp_results}
It can be seen in Figure~\ref{fig:integral_hist_scaled} that between \SI{25}~and \SI{150}{\countsamples} 
the excess of the \laserdata~spectrum above the \blankdata~spectrum differs significantly from a Gaussian tail, 
indicating the presence of a distinct population of under-amplified photoelectrons with low integrated charge. 
The mean of the single photoelectron distribution, as estimated by the method described in this paper, 
is shown by the dashed line in Figure~\ref{fig:integral_hist_scaled}. 
As one would expect,  the presence of under-amplified photoelectrons with lower output charge pushes the estimated mean distinctly 
below the peak of the fully-amplified single photoelectron distribution. For the \pmt~and operating gain described above, 
the estimated mean of the entire single-photoelectron distribution, including under-amplified photoelectrons, 
is $\approxim 80\%$ of the peak of the fully-amplified single-photoelectron distribution.   

\begin{figure}[t!]
\begin{center}
\ifcolorfigs
\includegraphics[width=\columnwidth]{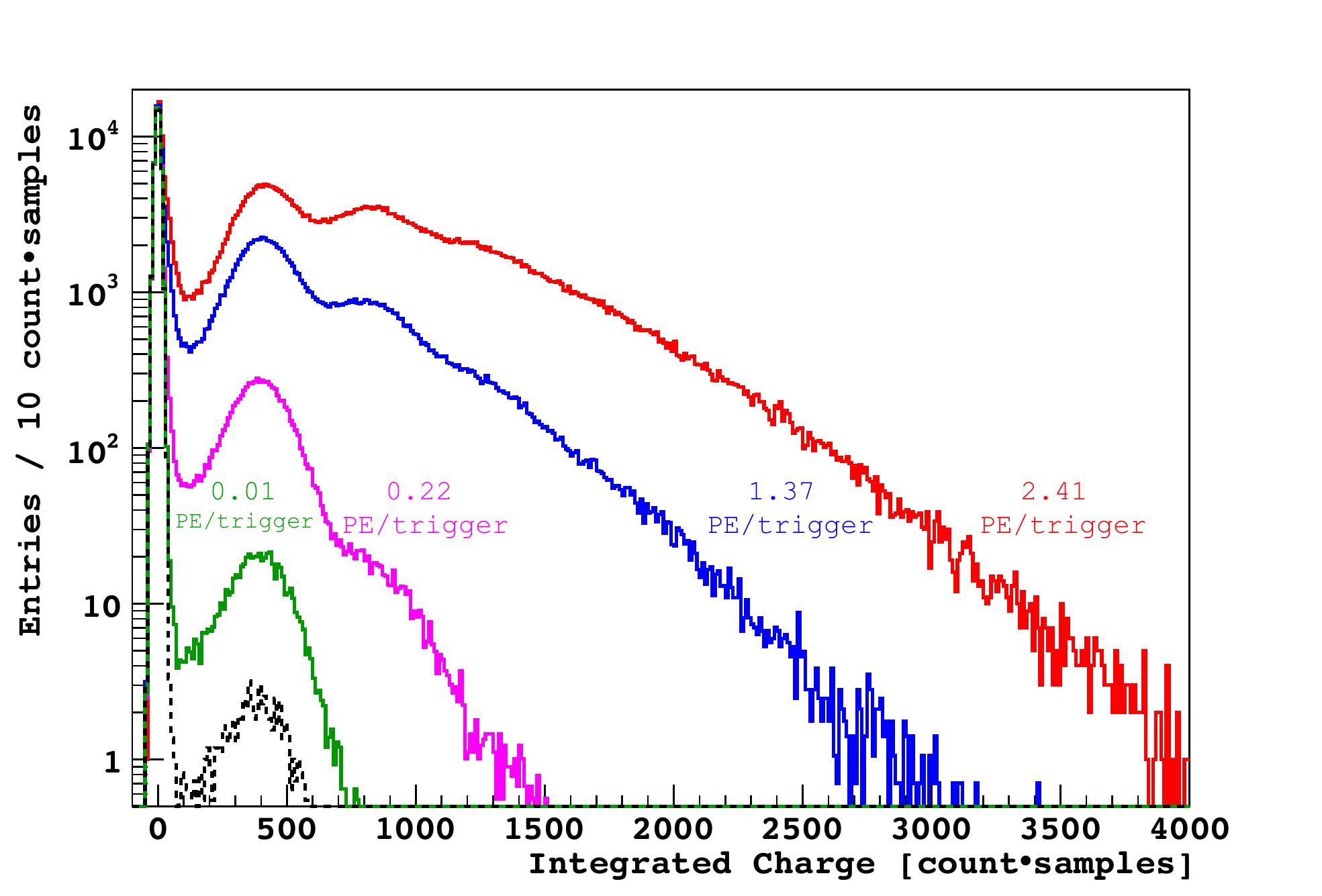}
\else
\includegraphics[width=\columnwidth]{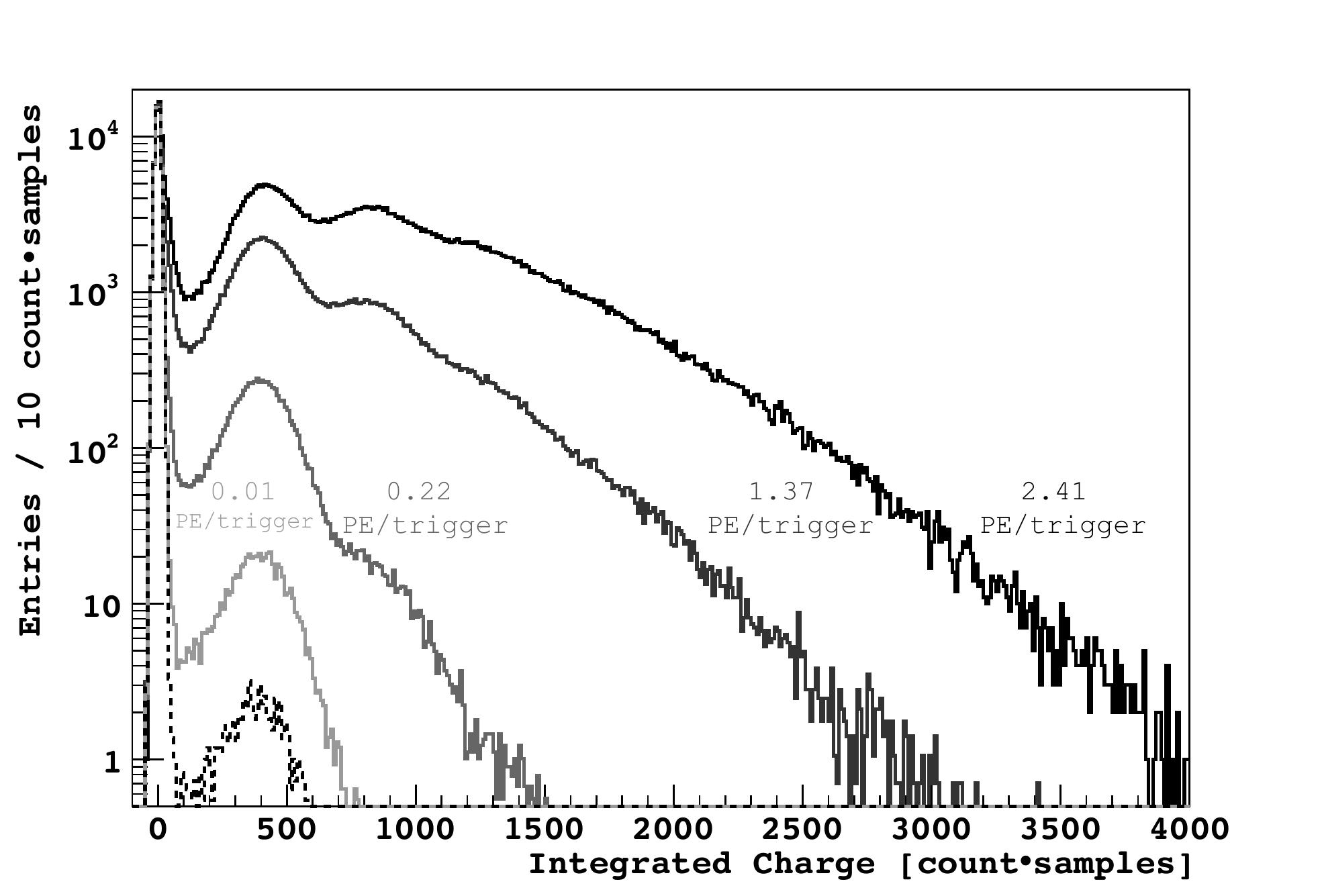}
\fi
\caption{Integrated charge spectra for \laserdata~data sets acquired at a voltage difference of 1700~V with different light intensities (\estim{\occup} = 2.41, 1.37, 0.22 and 0.01 \si{\pepertrigger}), along with the corresponding \blankdata~data set shown with a dashed line for comparison. The spectra are all scaled to have the same number of \zerope~triggers using the estimated occupancy.}
\label{fig:var_occ_integral_hist_scaled}
\end{center}
\end{figure}

\begin{table*}[t!]
\begin{center}
\begin{tabular}{c|cccc}
\hline
Nominal Attenuation & Occupancy & SPE Mean &  SPE Std. Dev. & SPE Rel. Std. Dev. \\
$\eta$ & \estim{\occup} & \sampmean{\sperand} & $\sampstdv{\sperand}$ & $\dfrac{\sampstdv{\sperand}}{\sampmean{\sperand}}$ \\
\\[-3ex]
 & [\si{\pepertrigger}] & [\si{\countsamples}] & [\si{\countsamples}] &  \\
\hline
1E5 & $2.412 \pm 0.015$ & $321.6 \pm 2.1$  & $186.3 \pm 1.7$ & $0.579 \pm 0.009$\\
2E5 & $1.374 \pm 0.010$ & $317.1 \pm 2.3$  & $187.4 \pm 1.9$ & $0.591 \pm 0.010$\\
1E6 & $0.216 \pm 0.006$ & $314.4 \pm 9.4$  & $189.5 \pm 8.0$ & $0.603 \pm 0.044$\\
1E7 & $0.012 \pm 0.003$ & $388 \pm 90$& $139 \pm 30$ & $0.36 \pm 0.13$\\
\hline
\end{tabular}
\end{center}
\caption{Results for the estimated occupancy, single photoelectron mean and standard deviation for data acquired at a fixed \pmt~voltage difference (1700 V), with different optical filters to vary the intensity of laser light.}
\label{tab:var_occ_results}
\end{table*}
\subsection{Dependence on Occupancy}
In order to study the robustness of the method with respect to intensity of laser light used, 
several data sets were taken with the \pmt~supplied at a fixed voltage, but with optical fibers providing different levels of attenuation of the laser light. 
The observed occupancy did not exactly scale with the nominal filter attenuation used because different numbers and combinations of filters were used to obtain the different light intensities and reflections can occur between the filters. 
Figure~\ref{fig:var_occ_integral_hist_scaled} shows the \laserdata~spectra we obtained, 
where the distributions have been normalized to have the same number of \zerope~triggers, based on the occupancy calculated as described in \sect~\ref{sec:par_estimation}.
The resulting estimates for the single photoelectron mean and standard deviation, $\sampstdv{\sperand} \equiv \sqrt{\sampvar{\sperand}}$, 
are shown in Table~\ref{tab:var_occ_results}. It can be seen that this method produces consistent results for the \spe~mean and variance, with statistical uncertainties below 3\% and 4\% respectively, 
over a range of occupancies that span from 0.2 to 2.4 \si{\pepertrigger}. 
At the lowest occupancy measured of $\occup = \SI{0.01}{\pepertrigger}$ the values obtained are still consistent with the other measurements, though the statistical uncertainties are too large for most applications. Better precision can be obtained simply by increasing the number of trigger samples acquired.
While our experimental data only extend up to an occupancy of 2.41 \si{\pepertrigger}, 
\eqn\eqref{eq:stat_unc_final} indicates that for the parameters of this experimental setup, and a suitable choice of threshold as described in \sect\ref{sec:thresh_choice},
the method has a statistical uncertainty of less than 3\% on the \spe~mean for occupancies spanning nearly two orders of magnitude from 0.1 to 9.5 \si{\pepertrigger}. 
This method is therefore especially useful for large detectors that contain an array of photomultiplier tubes, where a uniform illumination of the \pmts~may not be possible. 

\begin{figure}[t!]
\begin{center}
\ifcolorfigs
\includegraphics[width=\columnwidth]{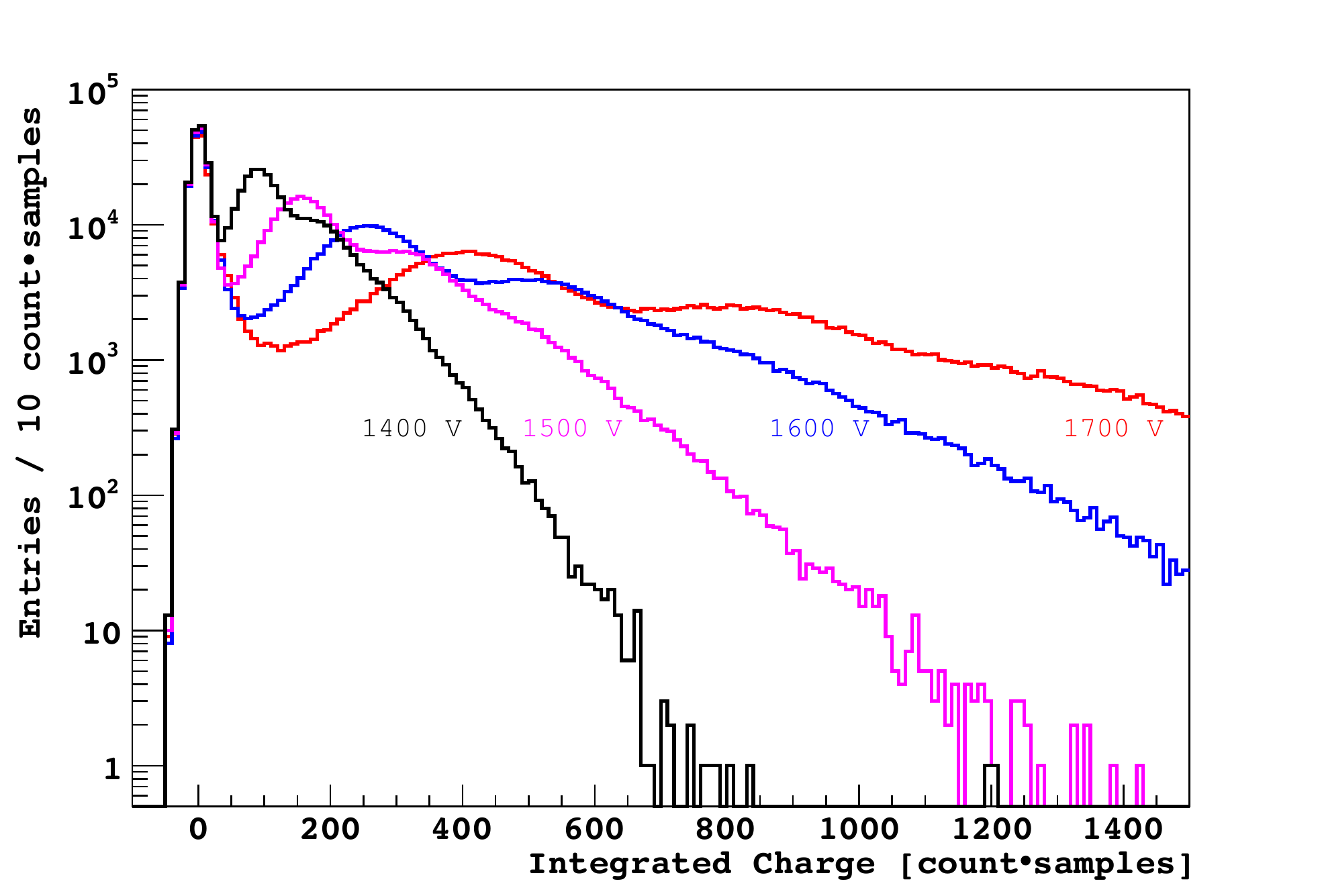}
\else
\includegraphics[width=\columnwidth]{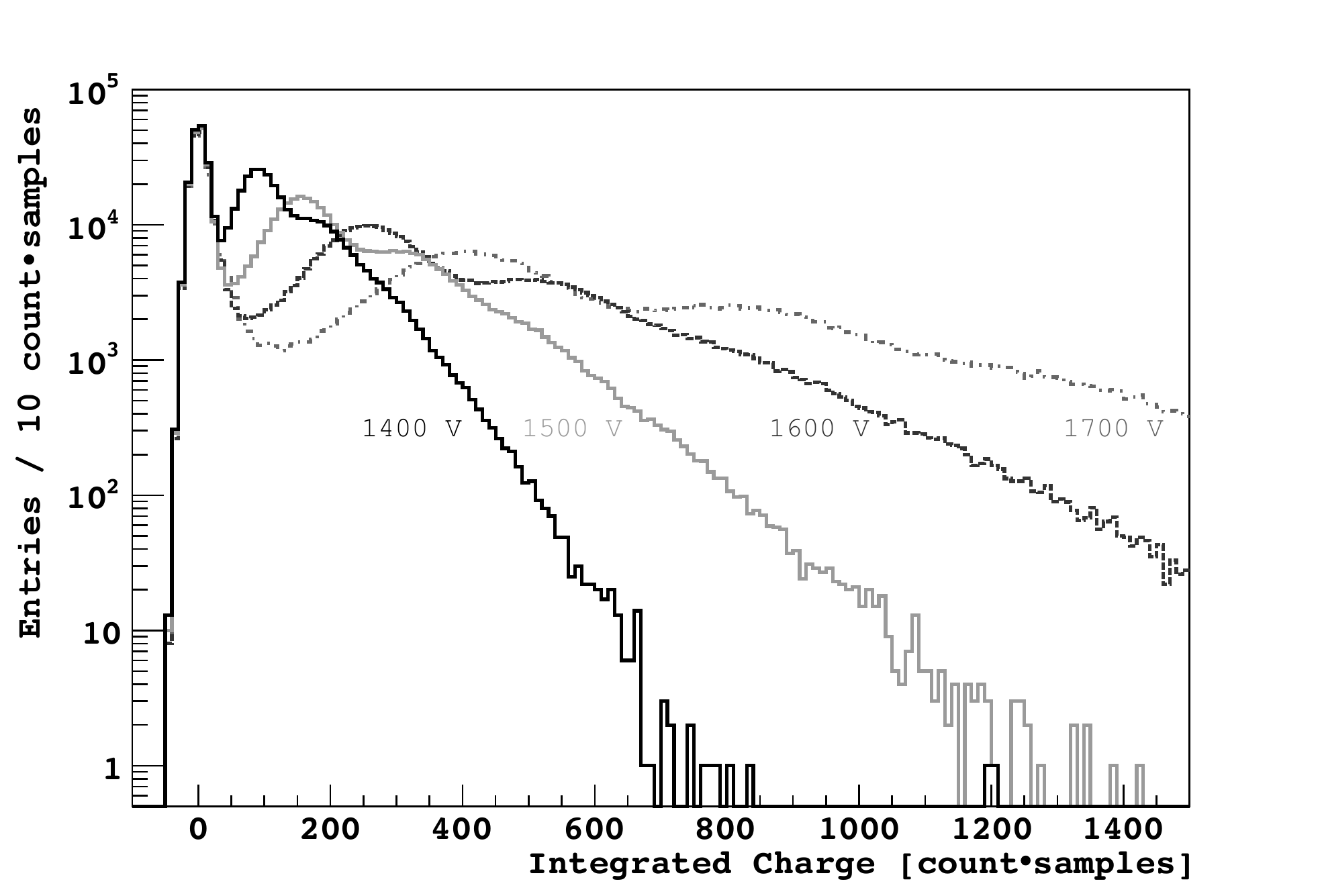}
\fi
\caption{Integrated charge spectra for \laserdata~data sets acquired at a fixed laser intensity 
(\estim{\occup} = 1.37~photoelectrons/trigger) with varying \pmt~voltage differences (1400 V, 1500 V, 1600 V, 1700 V).}
\label{fig:var_gain_integral_hist_scaled}
\end{center}
\end{figure}
\subsection{Dependence on Gain}
\label{subsec:exp_var_gain}
As described in Section~\ref{subsec:sys_unc}, the systematic uncertainty on the estimated \spe~mean is related to the fraction of single photoelectron triggers falling below the \thresholdcut~cut. One way the uncertainty can be reduced is by increasing the voltage applied to the \pmt, thereby increasing the gain. However, in some experimental setups it is often not possible to run \pmts~at high gain values due to dynamic range limitations or the emission of light from internal \pmt~structures \cite{Akimov20151}.
To study the performance of the method at lower gain, data sets were acquired at a constant laser intensity 
but varying voltage differences applied to the \pmt~(shown in Figure~\ref{fig:var_gain_integral_hist_scaled}). 

\begin{figure}[ht]
\begin{center}
\ifcolorfigs
\includegraphics[width=\columnwidth]{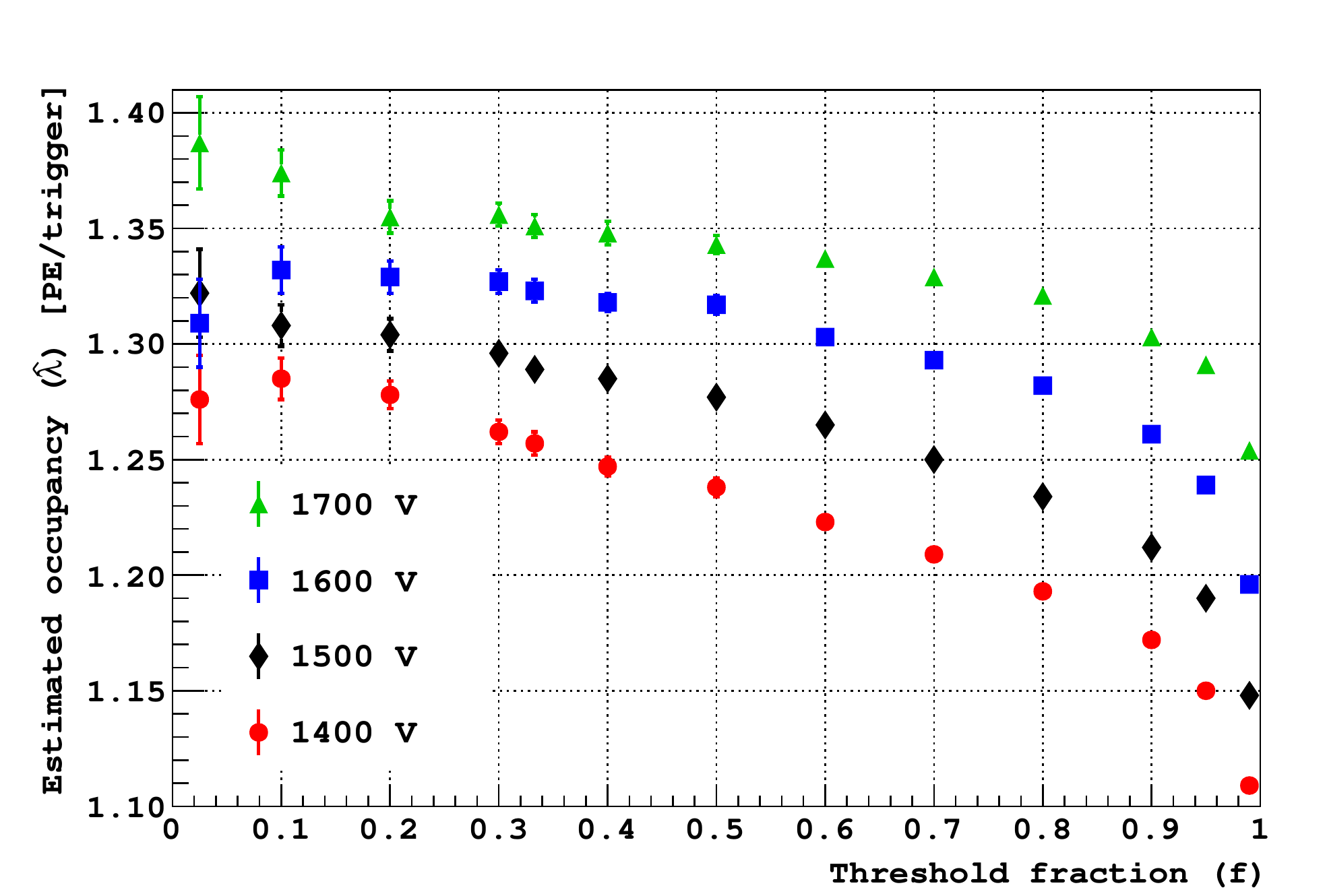}
\else
\includegraphics[width=\columnwidth]{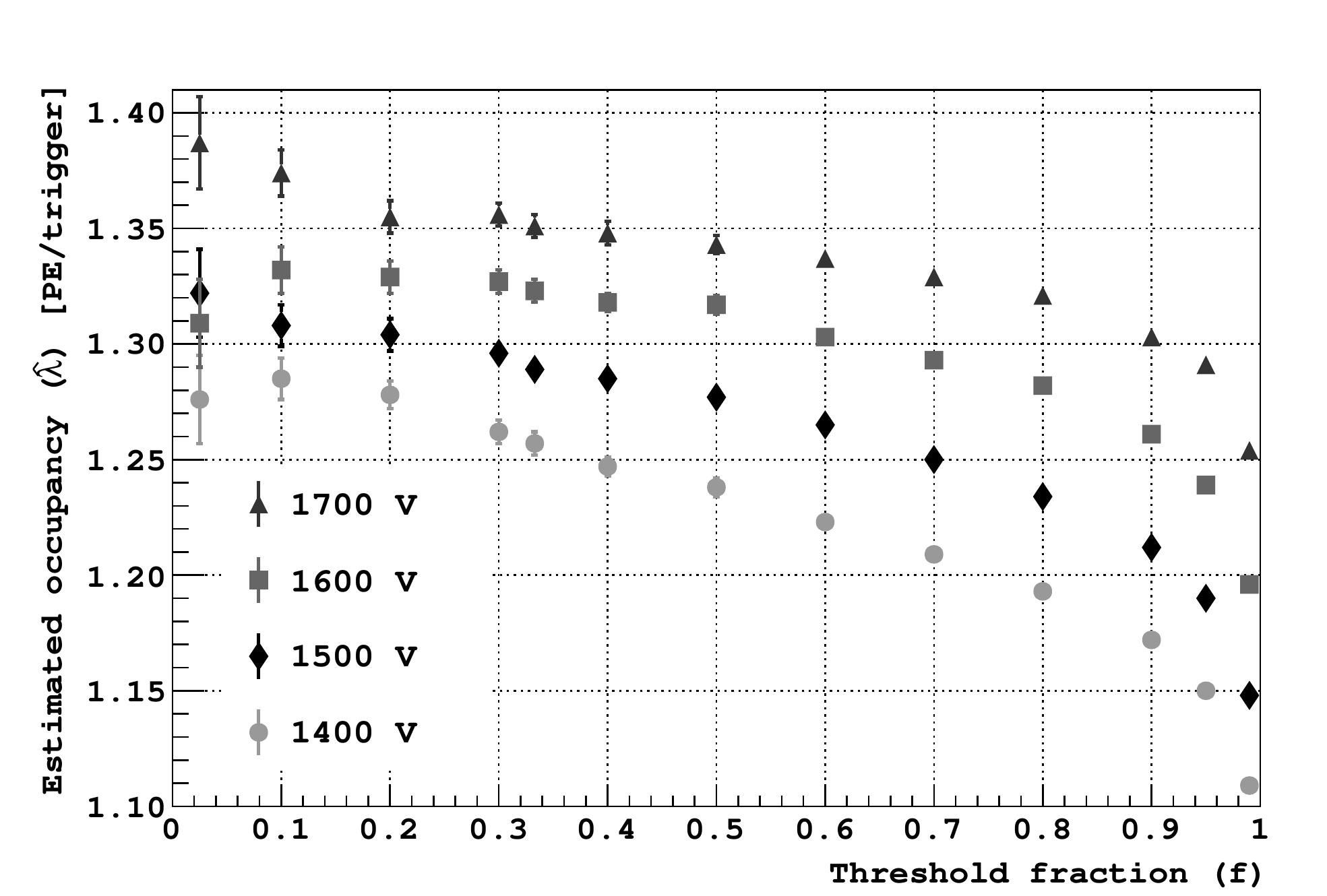}
\fi
\caption{Estimated occupancy for \laserdata~data sets acquired at a fixed light intensity with varying \pmt~voltage differences (1400 V, 1500 V, 1600 V, 1700 V) as a function of the fraction of single photoelectron triggers falling below the \thresholdcut~cut (\zeropefrac).}
\label{fig:occ_vs_f}
\end{center}
\end{figure}
Since the method described above uses a simple \thresholdcut~cut to estimate the occupancy, the overlapping of the noise 
and under-amplified photoelectrons makes it difficult to accurately estimate the occupancy and hence the single photoelectron mean and variance.
The estimated occupancy is shown in Figure~\ref{fig:occ_vs_f} as a function of the chosen threshold fraction \zeropefrac, for different \pmt~voltages.
It can be seen that choosing a lower threshold leads to an increase in the estimated occupancy. This is because as one lowers the threshold the fraction of underamplified photoelectrons falling below the threshold reduces and the systematic bias in the estimate of the occupancy is lower (see \sect\ref{subsec:sys_unc}). Similarly, running at higher \pmt~gains also decreases the fraction of underamplified photoelectrons falling below the threshold. The fractional variation in the estimated occupancy leads to roughly the same size fractional variation in the estimated \spe~mean, though in the opposite direction.

It should be noticed that even at a relatively low thresholds ($\zeropefrac \le 0.1$), the estimated occupancy decreases slightly as one lowers the \pmt~gain. While there may be some small loss of photomultiplier efficiency at lower voltages 
due to inefficient focussing in the dynode structure, as we will see from the simulation studies in \sect\ref{sec:monte_carlo}, the decrease in estimated occupancy (and correspondingly increase in estimated \spe~mean) is consistent with the expected systematic bias described in \sect\ref{sec:par_unc}. More sophisticated algorithms that rely on other parameters to estimate the occupancy will likely be less biased at low gains. 
\begin{figure}[t!]
\begin{center}
\ifcolorfigs
\includegraphics[width=\columnwidth]{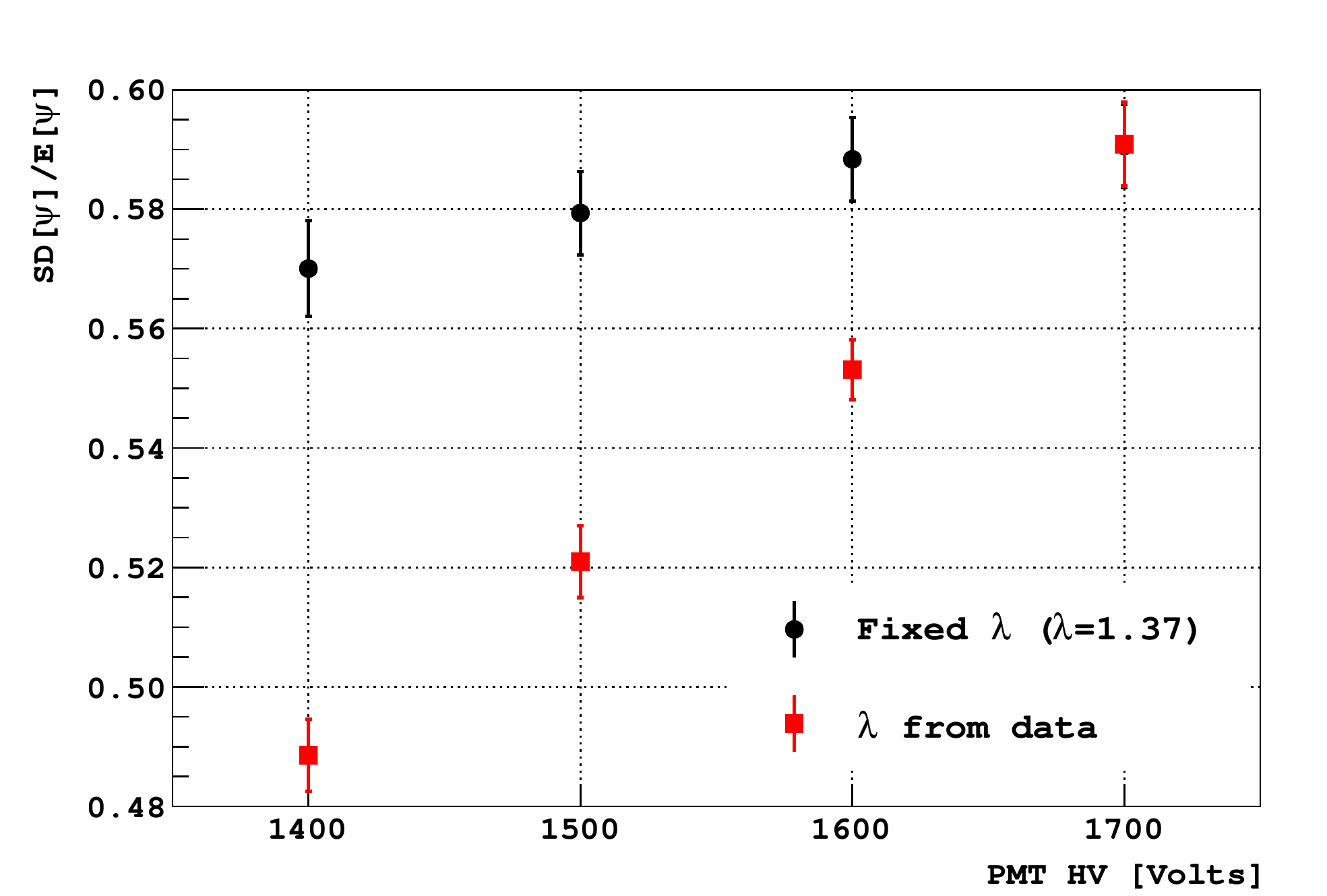}
\else
\includegraphics[width=\columnwidth]{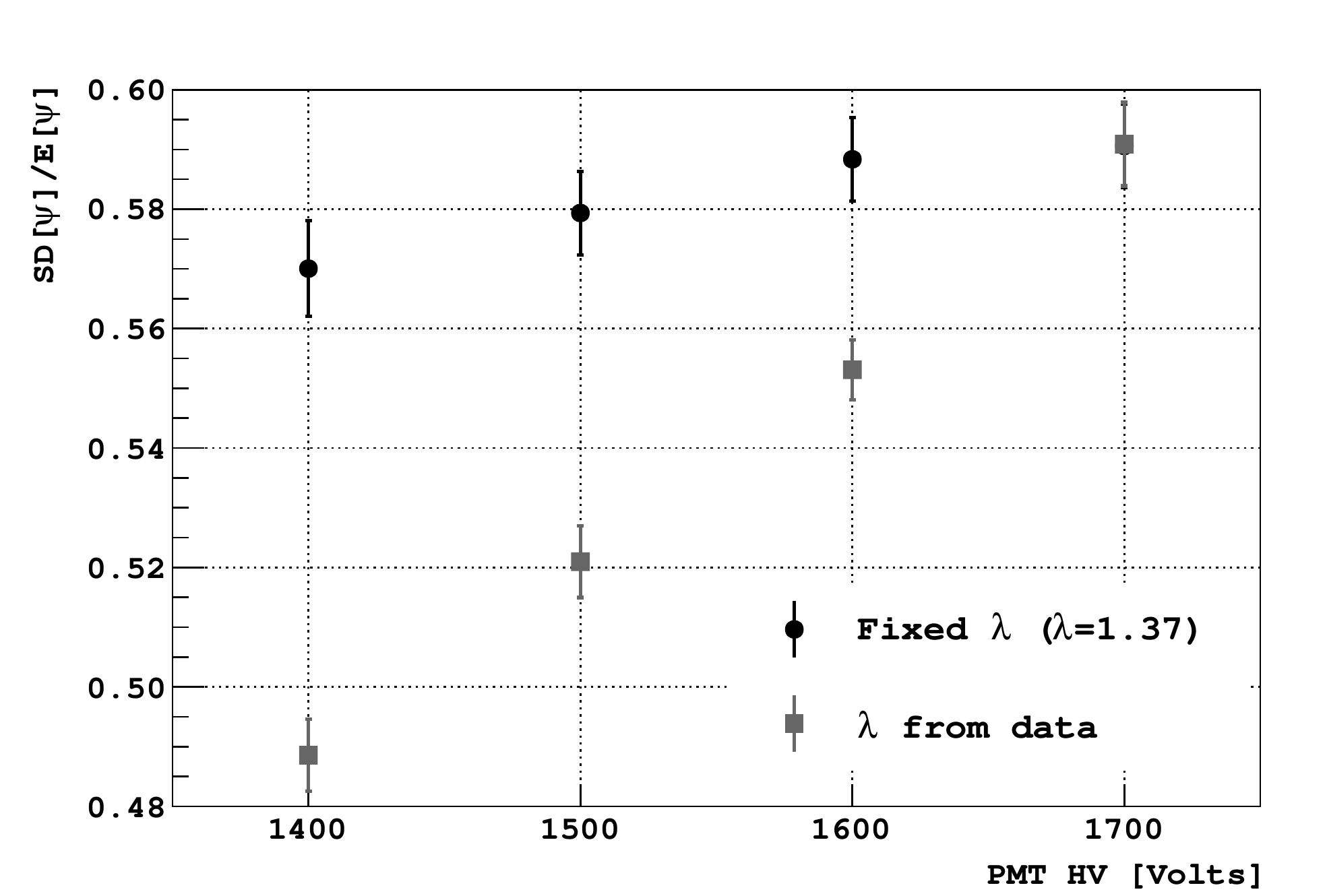}
\fi
\caption{Estimated single photoelectron relative standard deviation for data acquired 
at a fixed light intensity with varying \pmt~voltage differences. Circles: For all voltages the occupancy was set to the occupancy estimated at the highest gain and \zeropefrac = 0.1 (1.37 \si{\pepertrigger}). Squares: The occupancy was estimated independently at each of the corresponding voltages and \zeropefrac= 0.1.}
\label{fig:SDoverE_vs_PMTHV}
\end{center}
\end{figure}

Since the systematic bias increases as one lowers the gain, rather than trying to estimate the occupancy at each gain setting, one can fix the occupancy to the estimated value obtained at the highest gain, where the bias is least. We can compare the previous results with those using a fixed occupancy by comparing the estimated \spe~relative standard deviation, \sampstdv{\sperand}/\sampmean{\sperand}, which should remain roughly constant at all gain values (provided the \pmt~collection efficiency remains constant). Figure~\ref{fig:SDoverE_vs_PMTHV} shows that when fixing the occupancy to the value obtained at 1700 V the relative standard deviation is constant within 3\%, as compared to a 17\% variation when one uses the estimated occupancy obtained at each of the corresponding \pmt~voltages. Thus, for applications in which the photomultiplier is required to be calibrated at low gain, it is recommended to estimate the occupancy by temporarily running the PMT at a higher gain, while keeping the laser intensity constant. This value of the occupancy, estimated at higher gain, can then be used to calculate the single photoelectron mean and variance at the desired lower operating gain. If it is not possible to temporarily increase the \pmt~gain, the systematic bias in the estimated single photoelectron moments can be evaluated using simulations, as discussed in the following section.
\section{Simulation}
\label{sec:monte_carlo}
In order to verify that the single photoelectron calibration method described in this paper works accurately not only for the 
specific photomultiplier and conditions studied in the experimental setup, but also for different single photoelectron spectra, light intensities, and \pmt~gain, a Monte Carlo generator was written to simulate laser-induced PMT pulses and overlay them on background waveforms acquired during \blankdata~data sets. The simulated events were then processed in the same manner as the experimental data and the estimated single photoelectron moments were compared to the simulated inputs as a function of the gain, occupancy, and shape of the single photoelectron spectrum. To understand the systematic bias of the method we compared the mean value of the estimated moments, averaged over a large number of trials, to the simulated input; the standard deviation of the estimated moments was taken as the statistical uncertainty of the method.
\begin{figure}[t]
\begin{center}
\ifcolorfigs
\includegraphics[width=\columnwidth]{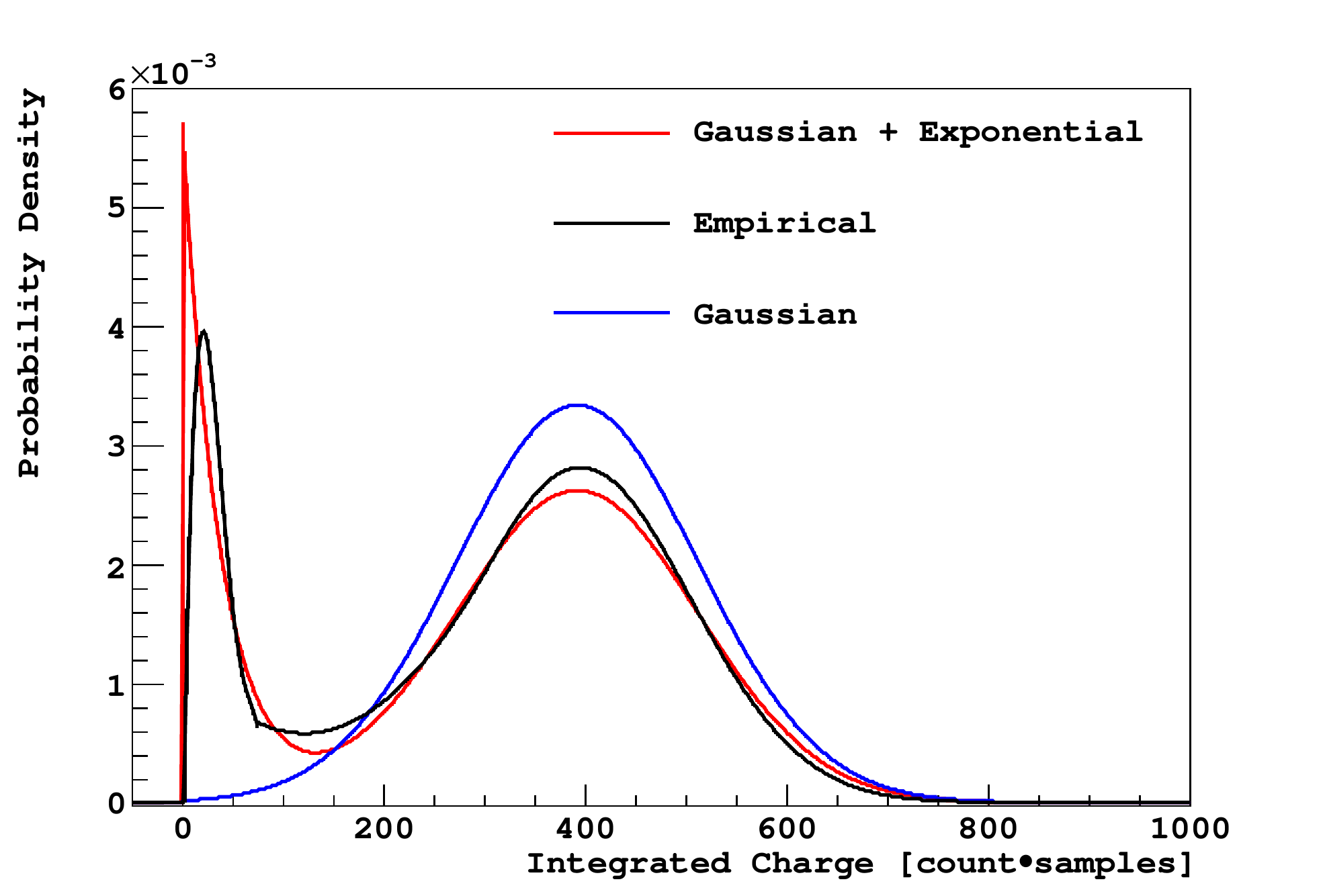}
\else
\includegraphics[width=\columnwidth]{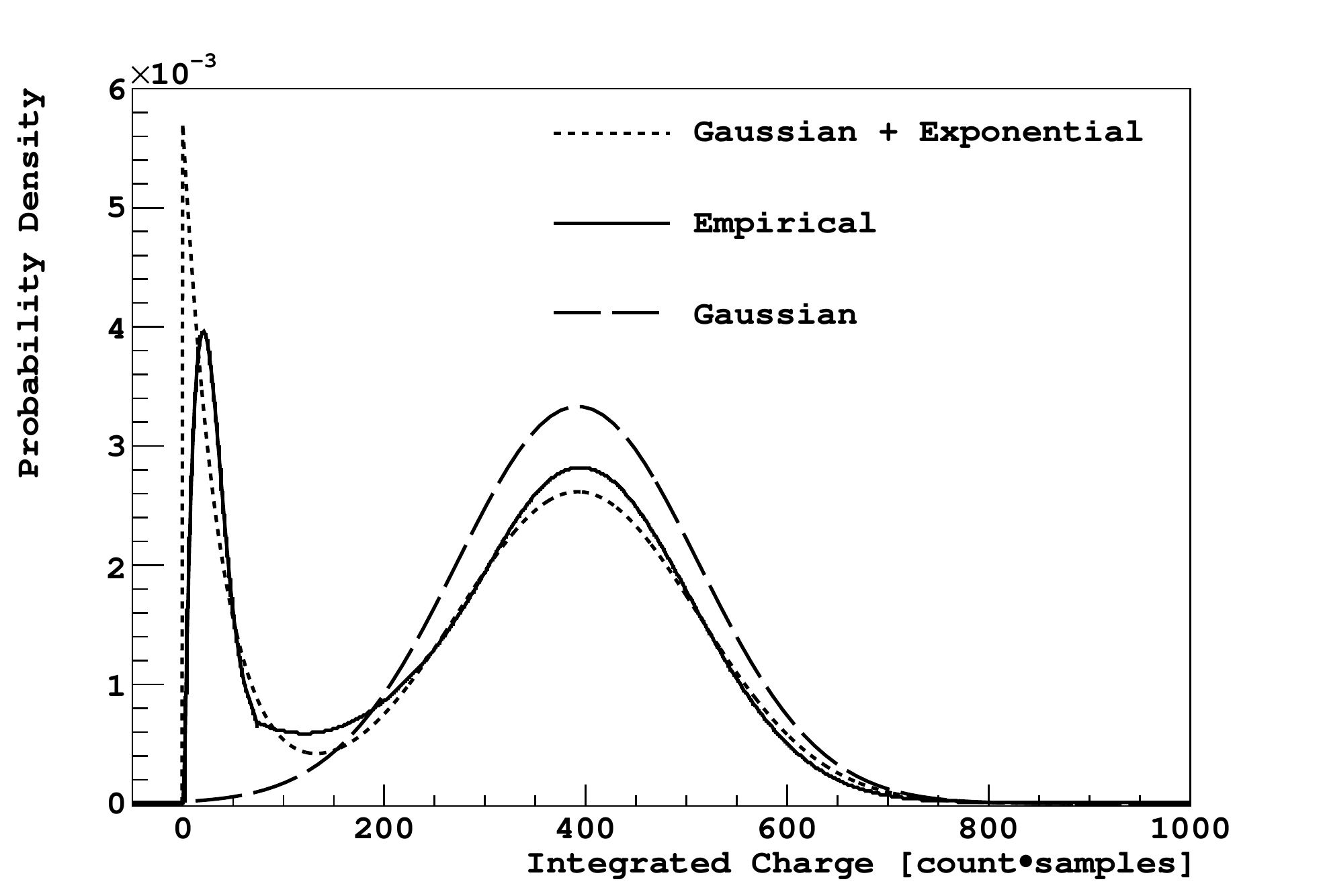}
\fi
\caption{Different spectral shapes used as single photoelectron distributions in the Monte Carlo simulations. 
The empirical distribution is a combination of a Gaussian distribution and an under-amplified component whose shape was obtained from 
the difference of the experimentally measured \laserdata~and \blankdata~spectra.}
\label{fig:spe_shapes}
\end{center}
\end{figure}
The simulation of each event in a given configuration begins by drawing a random number of photoelectrons from a Poisson distribution with a fixed mean corresponding to the desired occupancy. The integrated charge corresponding to each photoelectron was then independently drawn from a \spe~spectrum. 
Since the true shape of the \spe~charge spectrum is not known, three different approximations of the \spe~spectra were studied. The first spectrum was approximated from the experimental data, combining a Gaussian peak (representing the fully amplified photoelectron distribution) with an under-amplified distribution that was obtained by subtracting the scaled \blankdata~spectrum from the \laserdata~spectrum acquired at the highest gain and occupancy. This empirically derived \spe~spectrum, shown in Figure~\ref{fig:spe_shapes}, displays a prominent peak at low charge values, very similar to the shapes obtained in other experimental setups for a variety of different photomultiplier tubes \cite{Wright2010, Haas201106,ChirikovZorin2001310}. It should be noted that though this empirically derived shape serves as a good approximation of the true \spe~spectrum for the purposes of these simulation studies, it is not entirely accurate because the data from which it was derived included contributions from electronics noise present in each trigger as well as multiple photoelectrons. 
We considered two other shapes as potential extreme cases: a simple Gaussian truncated at zero, representative of a single photoelectron distribution without any contribution from under-amplified photoelectrons, and a Gaussian with an under-amplified distribution that rises exponentially at low charge values \cite{Dossi2000623}. For the latter two distributions, shown in Figure~\ref{fig:spe_shapes}, the shape of the spectra were tuned to try and match our experimental data as well as possible. In order to simulate different gains, 
the spectra were linearly scaled such that the peak of the Gaussian matched the experimental data at each \pmt~high voltage value.
\begin{figure*}[t!]
\begin{center}
\ifcolorfigs
\includegraphics[width=0.49\textwidth]{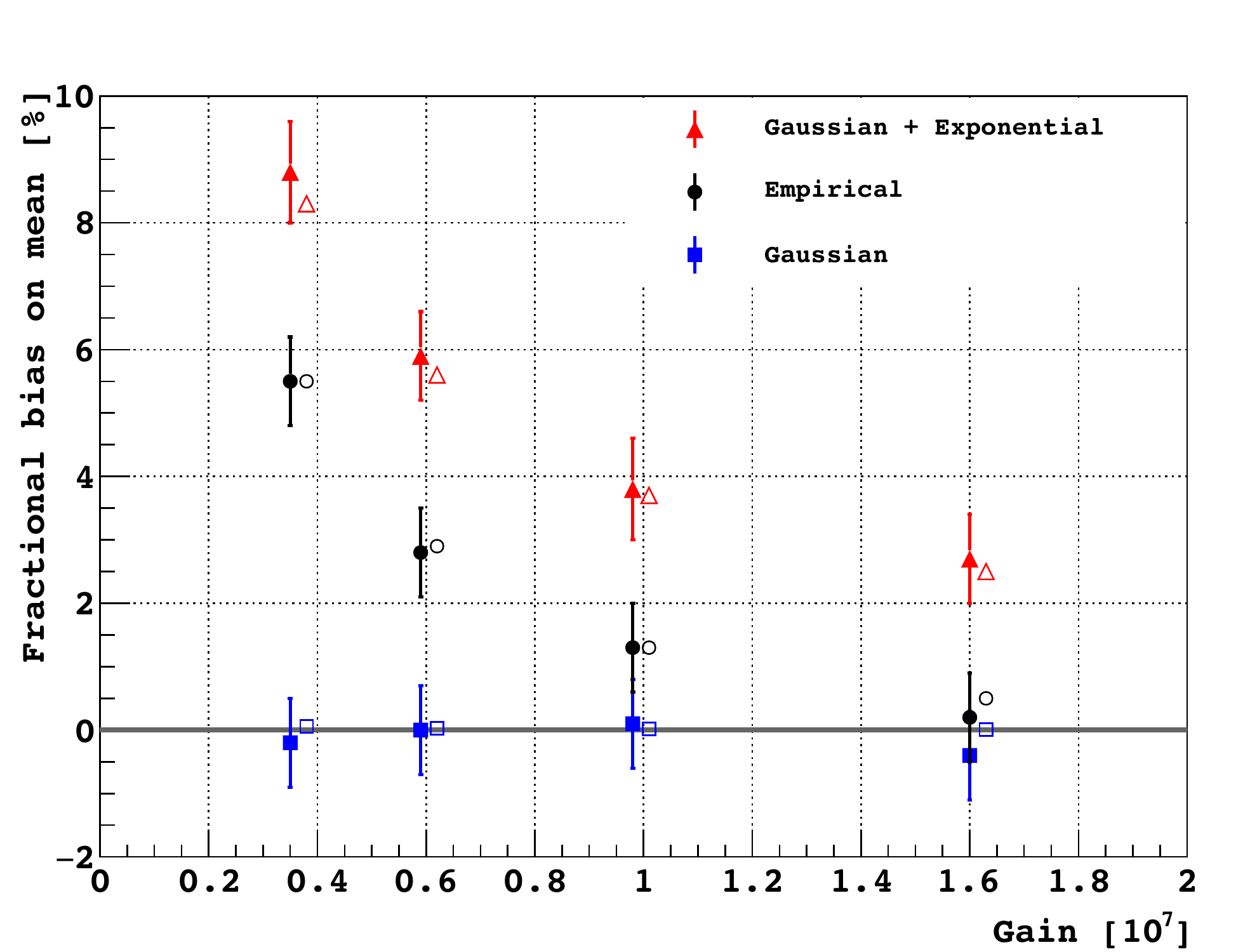}
\includegraphics[width=0.49\textwidth]{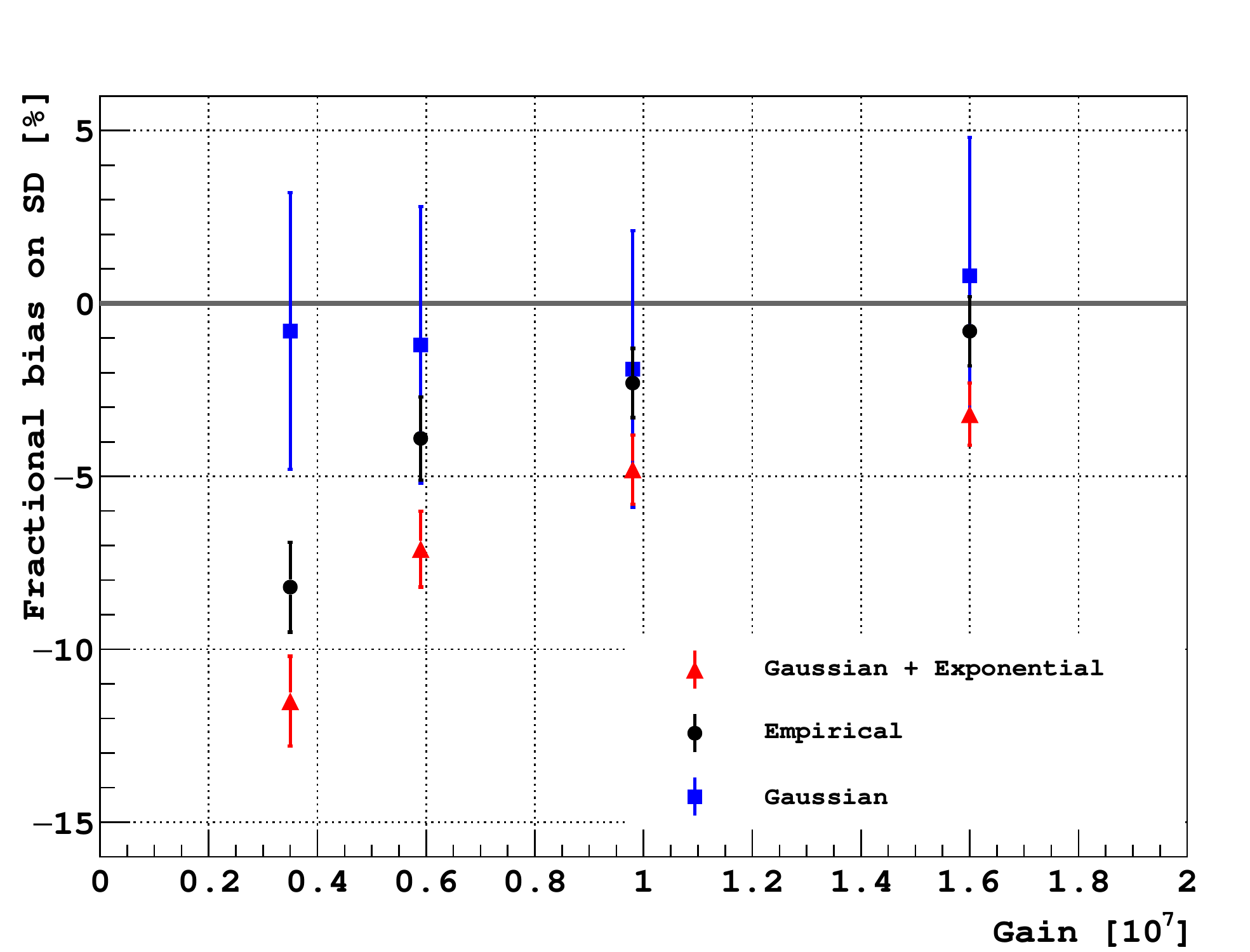}
\else
\includegraphics[width=0.49\textwidth]{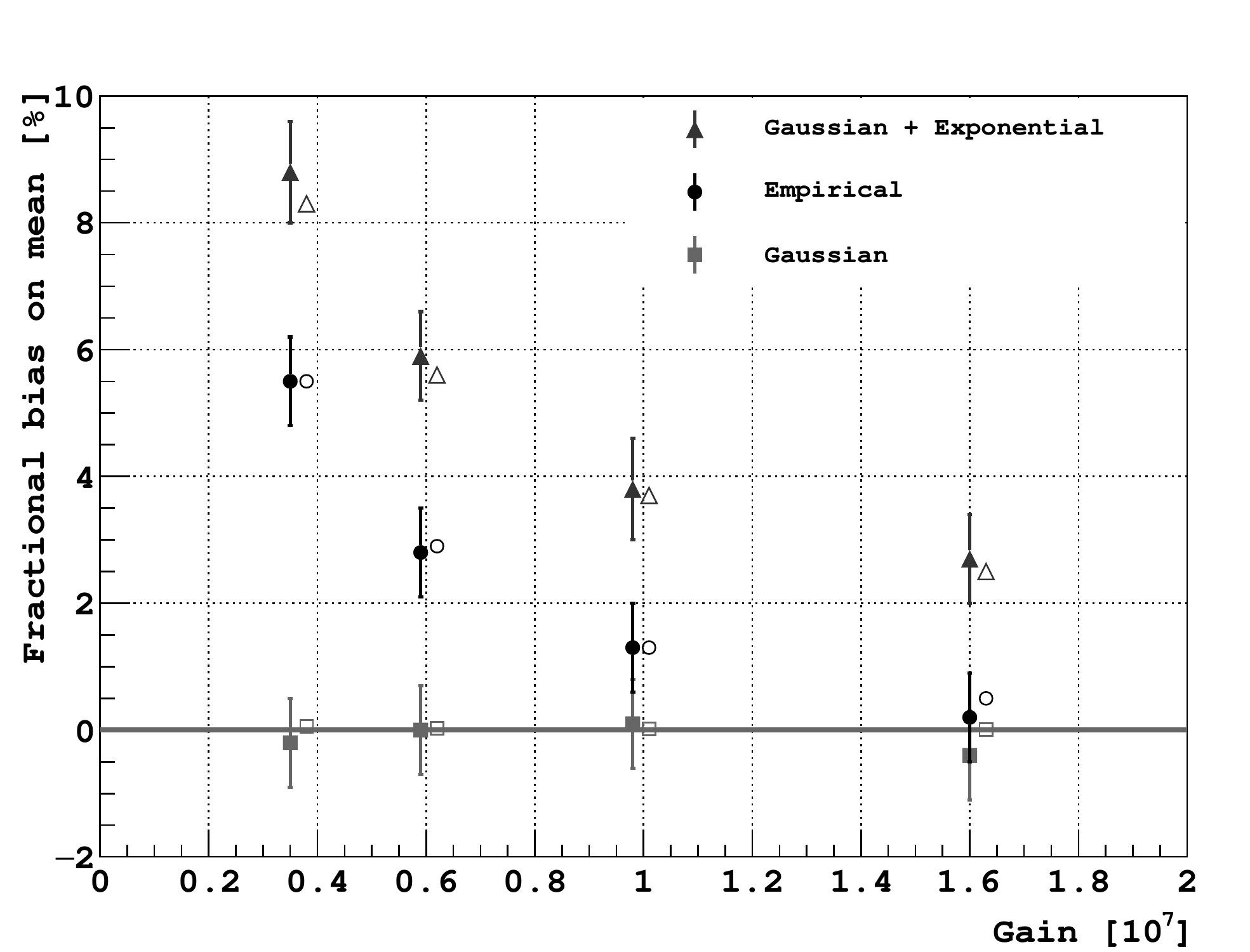}
\includegraphics[width=0.49\textwidth]{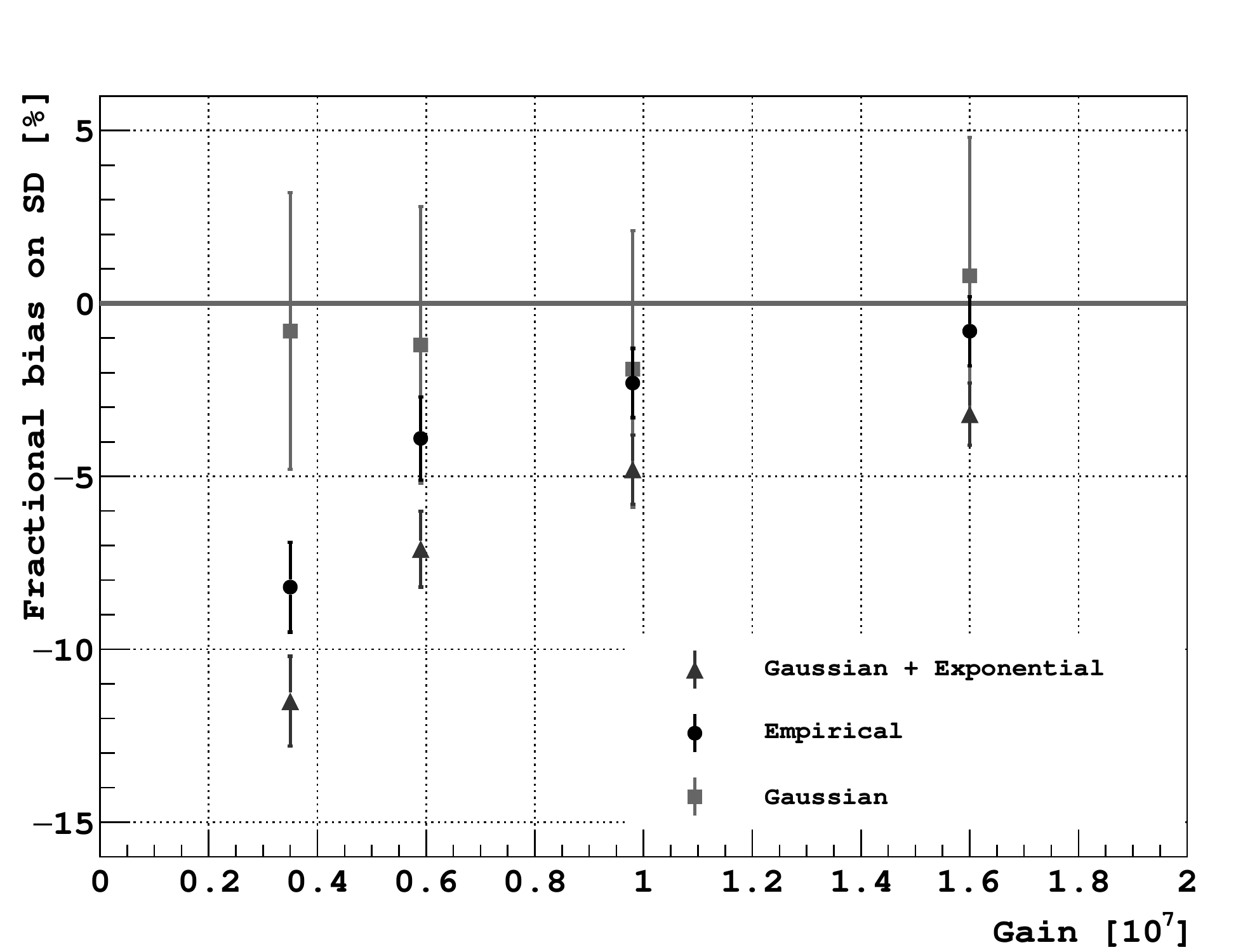}
\fi
\caption{Simulation results for a  fixed occupancy (1.37 \si{\pepertrigger}), with varying gain.
Left: Fractional bias in the estimated single photoelectron mean $\left(\frac{\mean{\sampmean{\psi}} - \simmean{\psi}}{\simmean{\psi}}\right)$.
Right: Fractional bias in the estimated standard deviation $\left(\frac{\mean{\sampstdv{\psi}} - \simstdv{\psi}}{\simstdv{\psi}}\right)$.
The filled markers indicate the fractional bias of the model independent method for the different simulated single photoelectron distributions, while the error bars depict the fractional statistical uncertainty of the method. 
The open markers on the left (horizontally displaced for clarity) indicate the analytic calculation of the fractional bias in the mean for each case (see \eqn\eqref{eq:sys_unc_final}).}
\label{fig:mc_var_gain}
\end{center}
\end{figure*}
\begin{figure*}[t!]
\begin{center}
\ifcolorfigs
\includegraphics[width=0.49\textwidth]{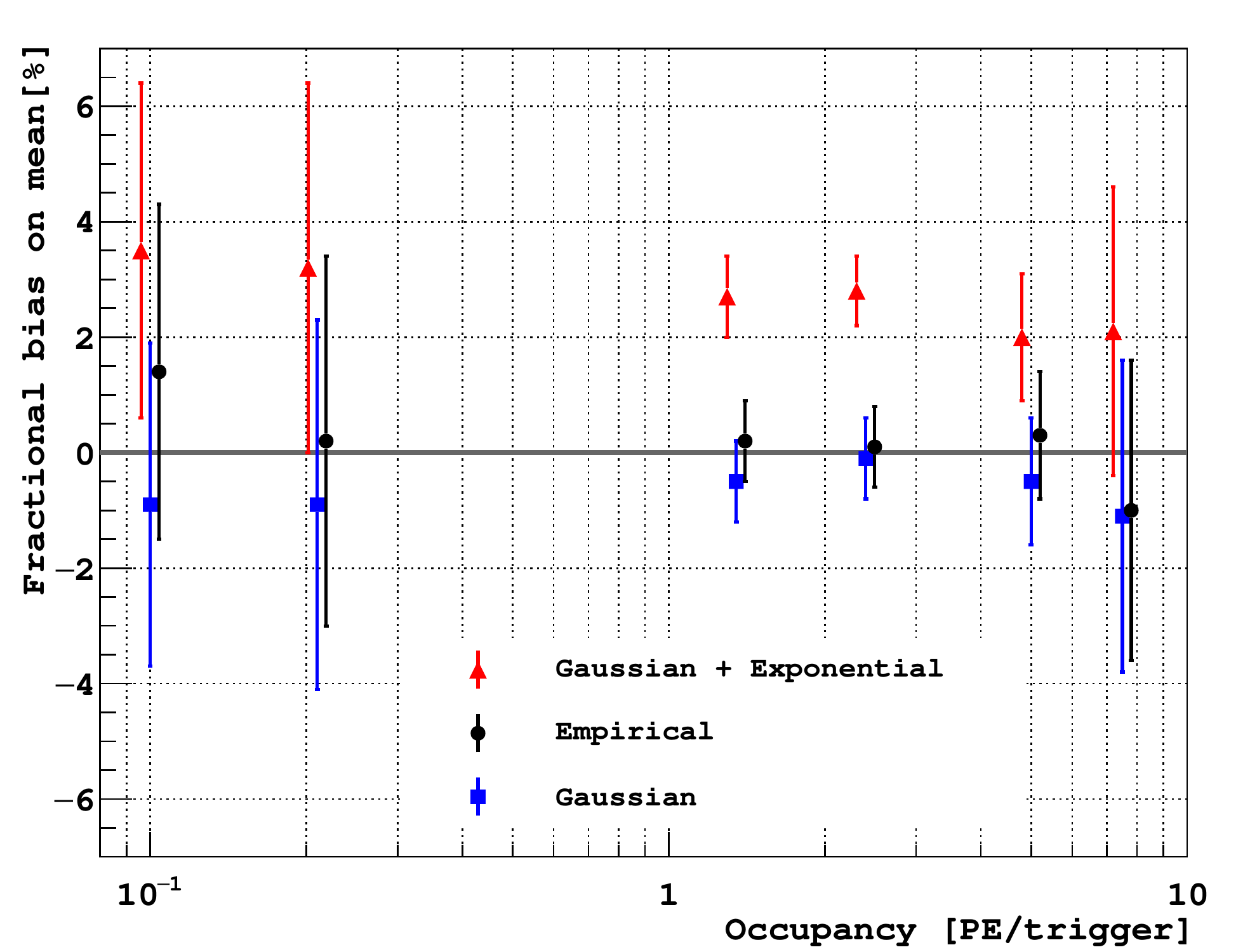}
\includegraphics[width=0.49\textwidth]{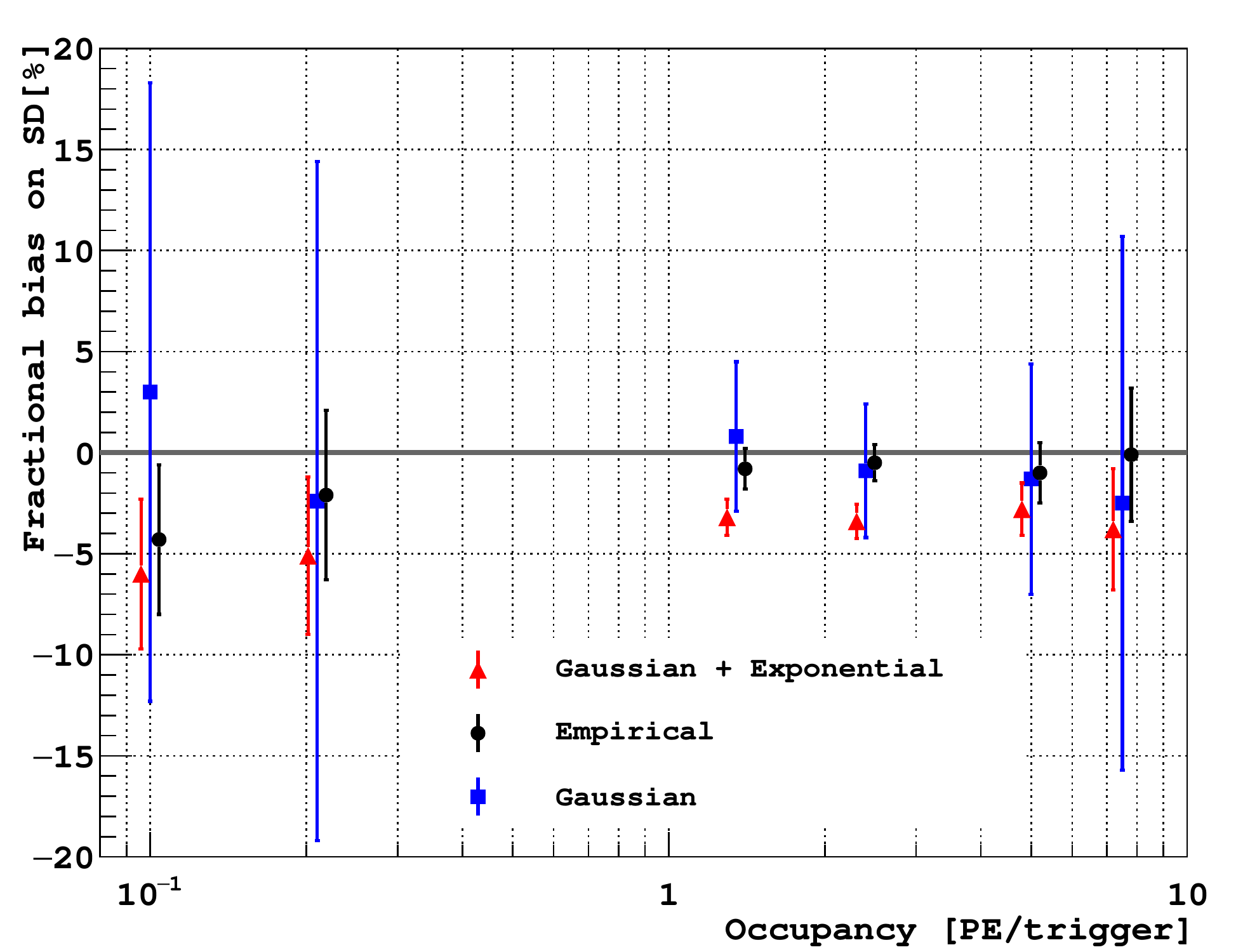}
\else
\includegraphics[width=0.49\textwidth]{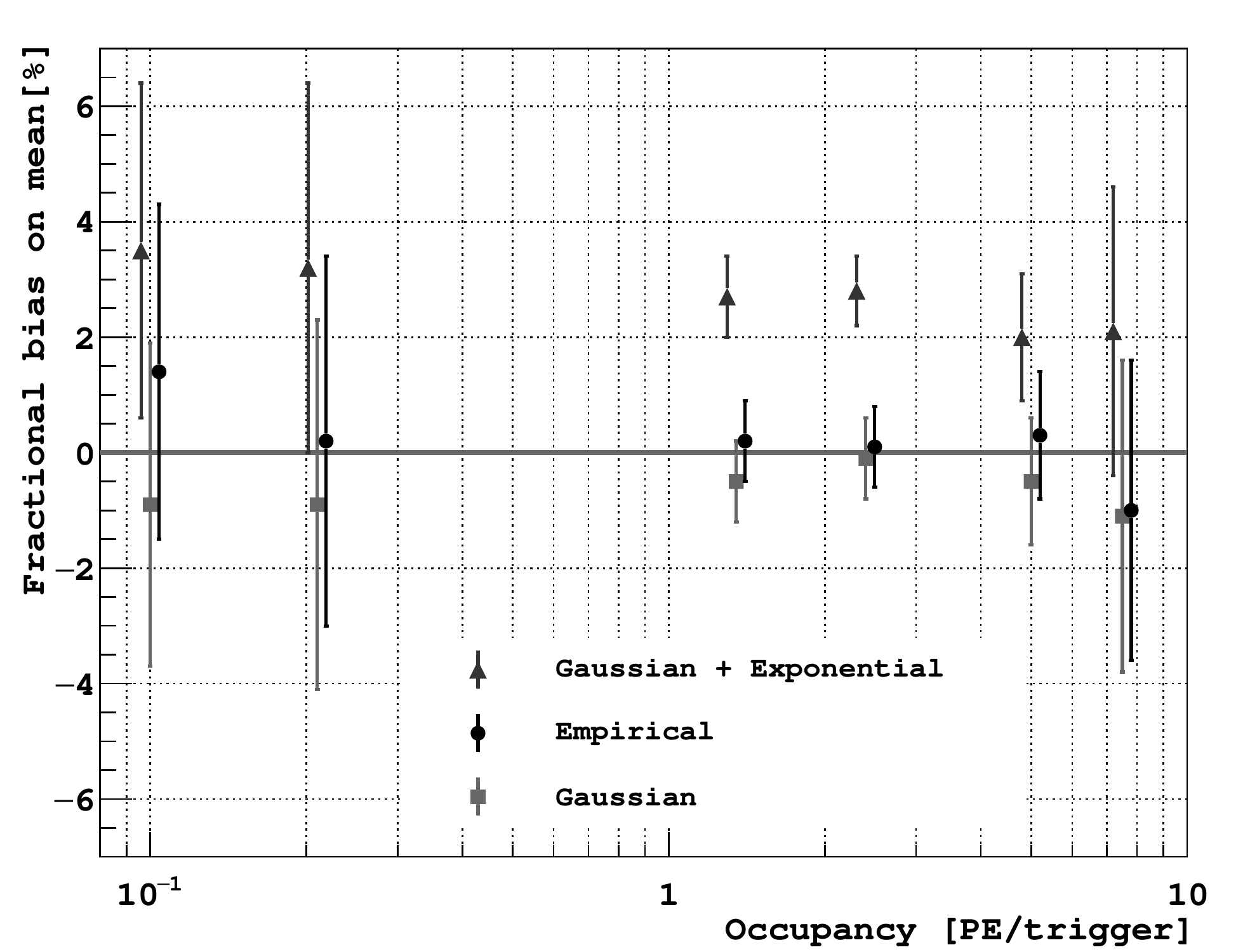}
\includegraphics[width=0.49\textwidth]{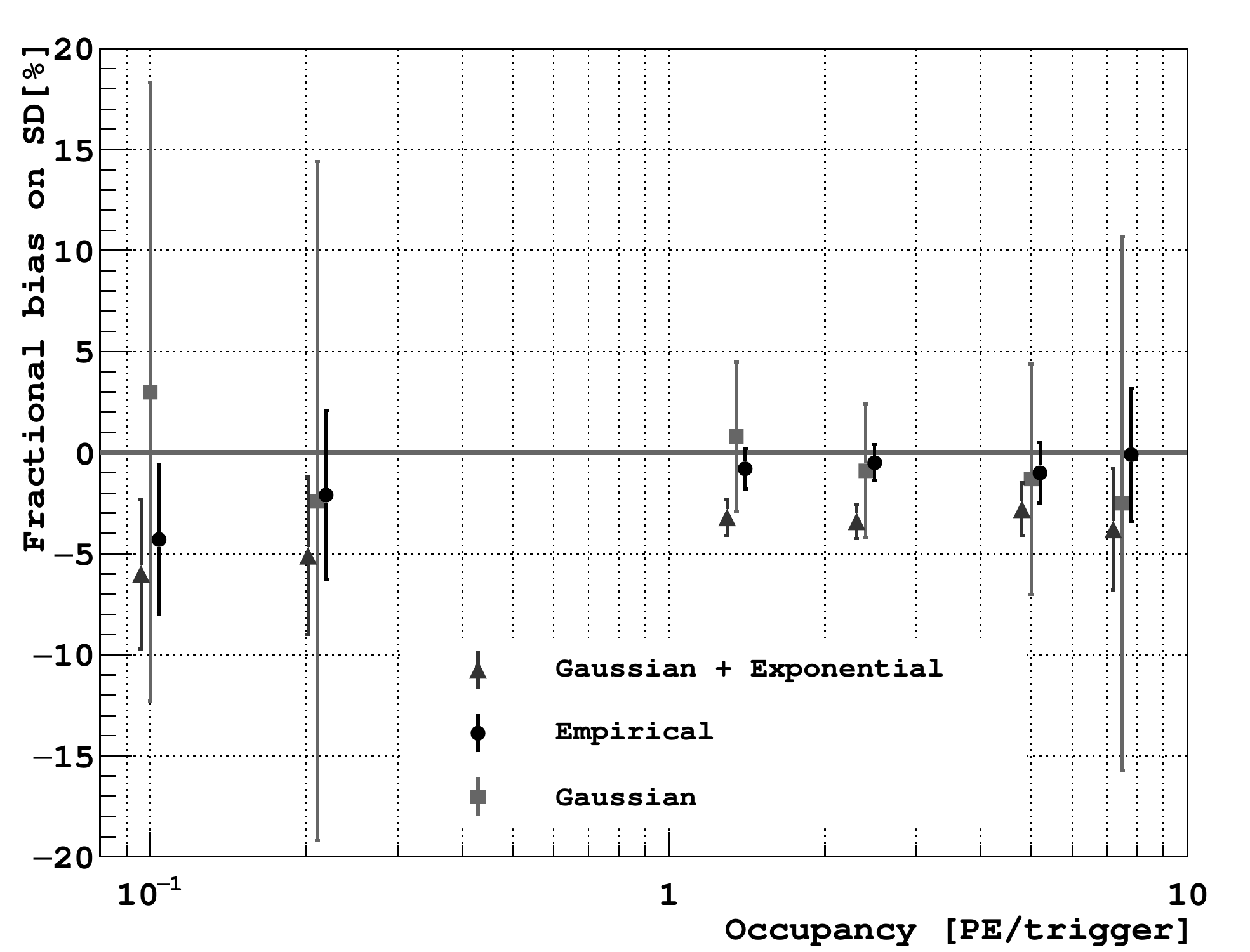}
\fi
\caption{Simulation results for a fixed gain (\pmtgainseventeen), with varying occupancy. Left: Fractional bias in the estimated single photoelectron mean $\left(\frac{\mean{\sampmean{\psi}} - \simmean{\psi}}{\simmean{\psi}}\right)$. Right: Fractional bias in the estimated standard deviation $\left(\frac{\mean{\sampstdv{\psi}} - \simstdv{\psi}}{\simstdv{\psi}}\right)$. The markers indicate the fractional bias of the model independent method for the different simulated single photoelectron distributions (horizontally displaced for clarity), while the error bars depict the fractional statistical uncertainty of the method.}
\label{fig:mc_var_occ}
\end{center}
\end{figure*} 
For each simulated photoelectron a waveform is generated based on measurements of the single photoelectron pulse shape. The integral of the pulse is scaled to equal the charge assigned to the photoelectron and the peak time is set to match the arrival time of the \pmt~signal in the experimental setup. The pulses corresponding to all the photoelectrons in a given event are then summed together and overlaid on top of a waveform acquired during a \blankdata~data set taken in the same configuration, accounting for the discreteness of the ADC samples in time and amplitude. The use of the experimentally obtained waveforms from the \blankdata~data set allows us to accurately include all of the relevant effects such as noise, stray photoelectrons and dark current into the simulation. 
Each simulated waveform is processed identically to the experimental data in order to obtain the estimates of the \spe~mean and variance.
\subsection{Dependence on Gain}
Using the results of the simulation, we studied the effects of the systematic bias observed in the experimental data (\sect\ref{subsec:exp_var_gain}) for the three 
different single photoelectron spectra as a function of the \pmt~gain. At lower gains the bias is expected to increase 
due to the increasing fraction of single photoelectron triggers falling below the \thresholdcut~cut. 
In Figure~\ref{fig:mc_var_gain} we report the results of the simulation for single photoelectron gain settings corresponding 
to the \pmt~HV supply values used in the experimental setup, and an occupancy of 1.37 \si{\pepertrigger}. 
As expected, the systematic bias is larger at lower gains and for simulated \spe~spectra with a larger underamplified component. 
The consistency between the simulation results (closed circles) and the estimated systematic uncertainty in \eqn\eqref{eq:sys_unc_final} (open circles),
indicates that the analytical model derived in \append\ref{sec:bias} is accurate. 
For a truly Gaussian \spe~spectrum the estimates of the mean and variance of the \spe~response are unbiased at all gains, 
while for the empirical \spe~spectrum obtained from the experimental measurements of the R11410, the bias on the mean (standard deviation) 
ranges from 0\% to +5\% (-1\% to -8\%), depending on the gain. These values closely match the variations in the experimental data presented in \sect\ref{subsec:exp_var_gain}, suggesting that the simulation accurately represents the experimental data. It should be noted that in the worst case considered, 
with an exponentially increasing under-amplified spectrum and low \pmt~gain, the systematic bias in the single photoelectron mean and 
standard deviation of +8\% and -12\% respectively is significantly lower than the bias one would obtain by ignoring the under-amplified electrons and only fitting the Gaussian component (+20\% and -50\% respectively).
\subsection{Dependence on Occupancy}
In order to test the sensitivity of the method to the intensity of the laser light, we simulated datasets at various occupancies
for all three different single photoelectron spectra at a gain equivalent to the experimental data taken at 1700 V. 
The results are shown in Figure~\ref{fig:mc_var_occ}, where it can be seen that the estimates are consistent 
with the simulated single photoelectron moments (after accounting for the small systematic bias discussed above). 
This confirms the validity of the method for a wide range of \pmt~illumination. The larger fractional statistical uncertainty of the estimated \spe~standard deviation obtained with the Gaussian \spe~spectrum, as compared to the other spectra, is due to the larger ratio of the \spe~mean to variance (see \eqn\eqref{eq:var_stat_unc_final}).

\subsection{Comparison with fit methods}
\begin{figure*}[t!]
\begin{center}
\ifcolorfigs
\includegraphics[width=0.49\textwidth]{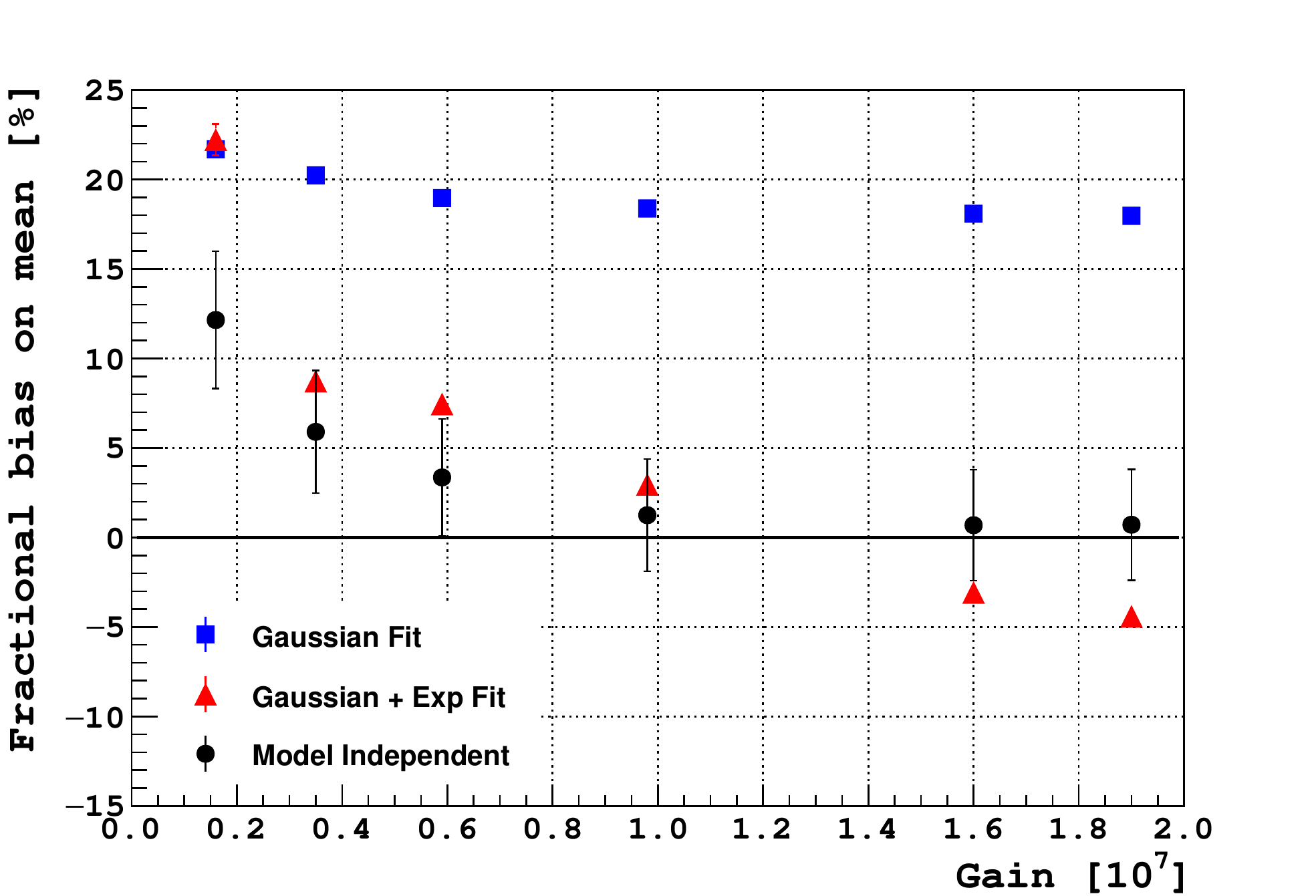}
\includegraphics[width=0.49\textwidth]{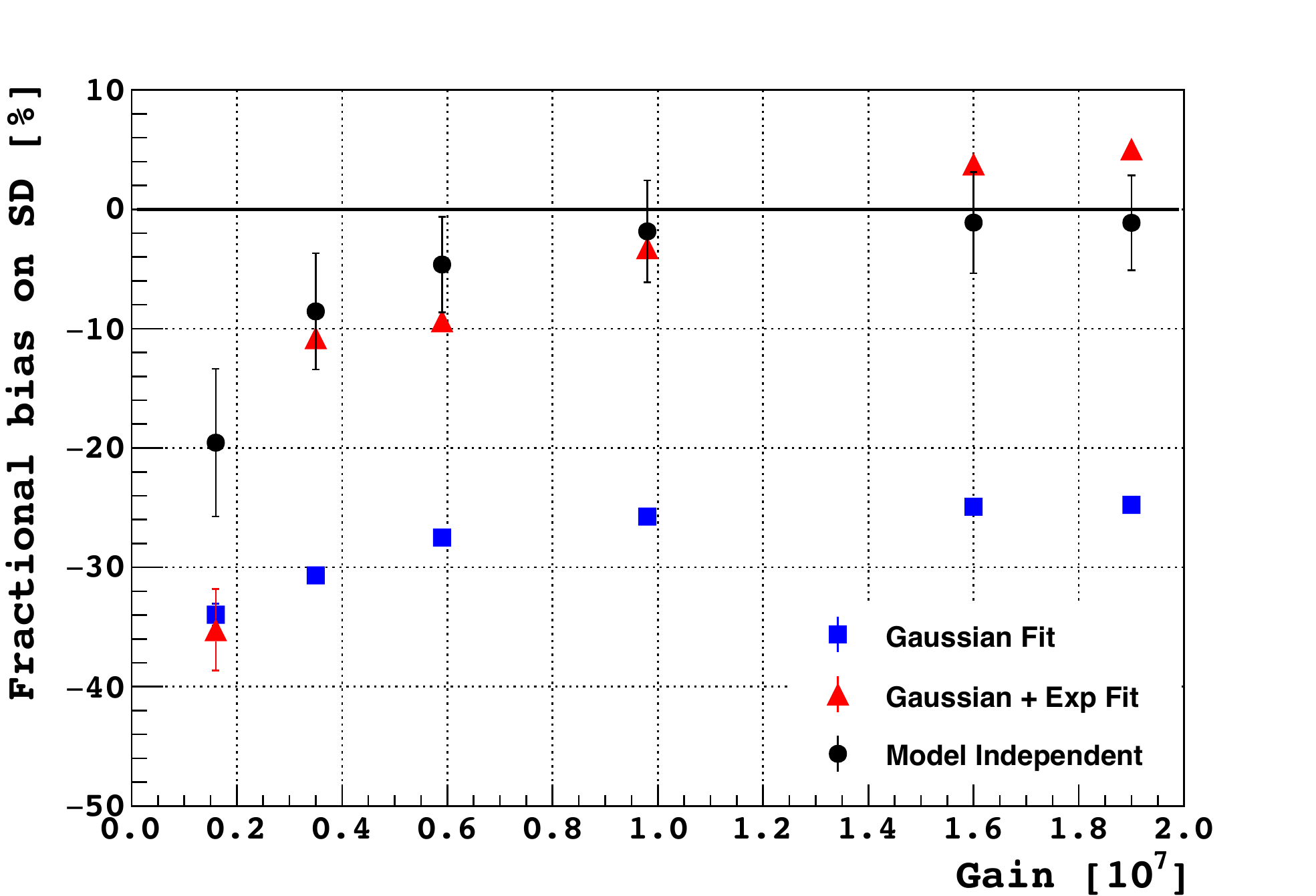}
\else
\includegraphics[width=0.49\textwidth]{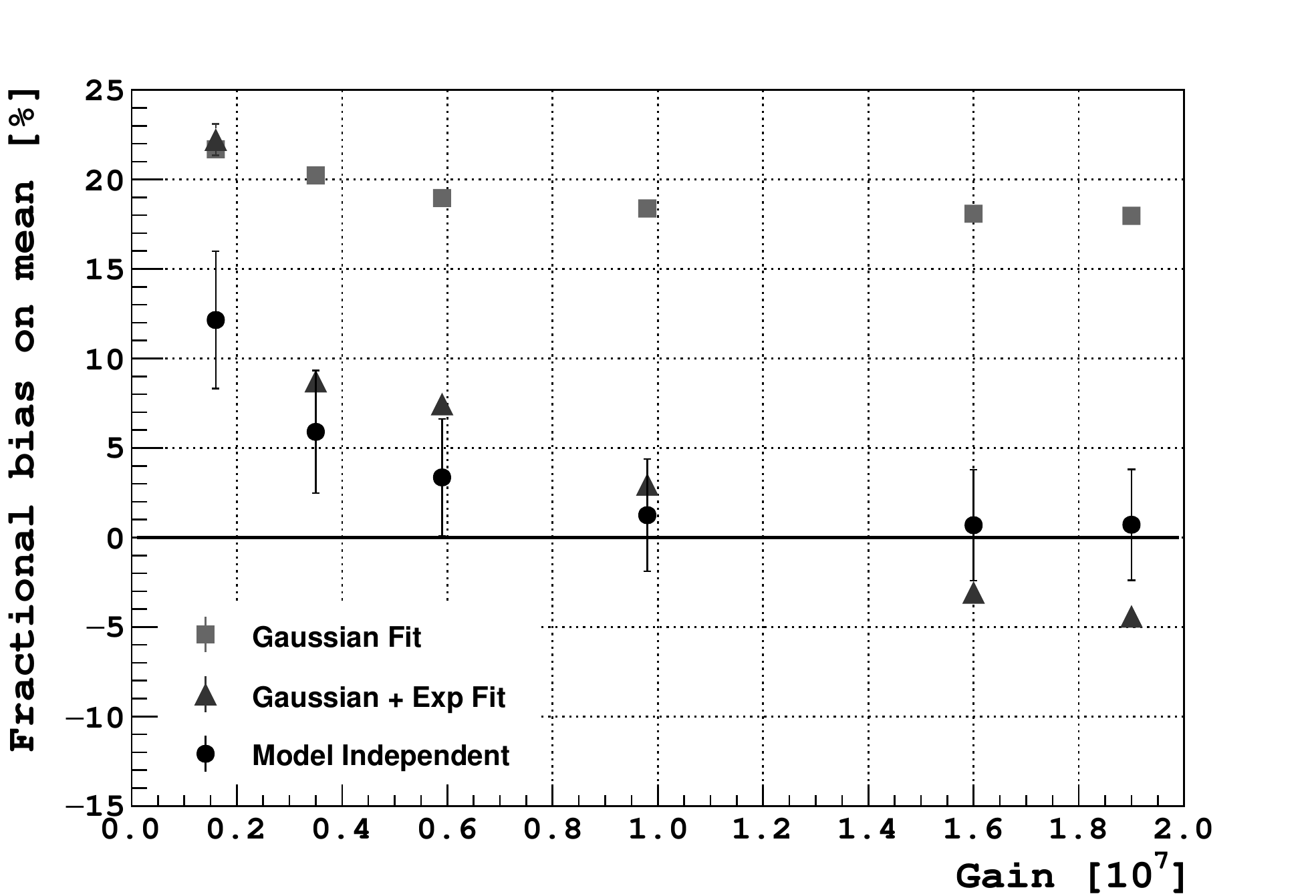}
\includegraphics[width=0.49\textwidth]{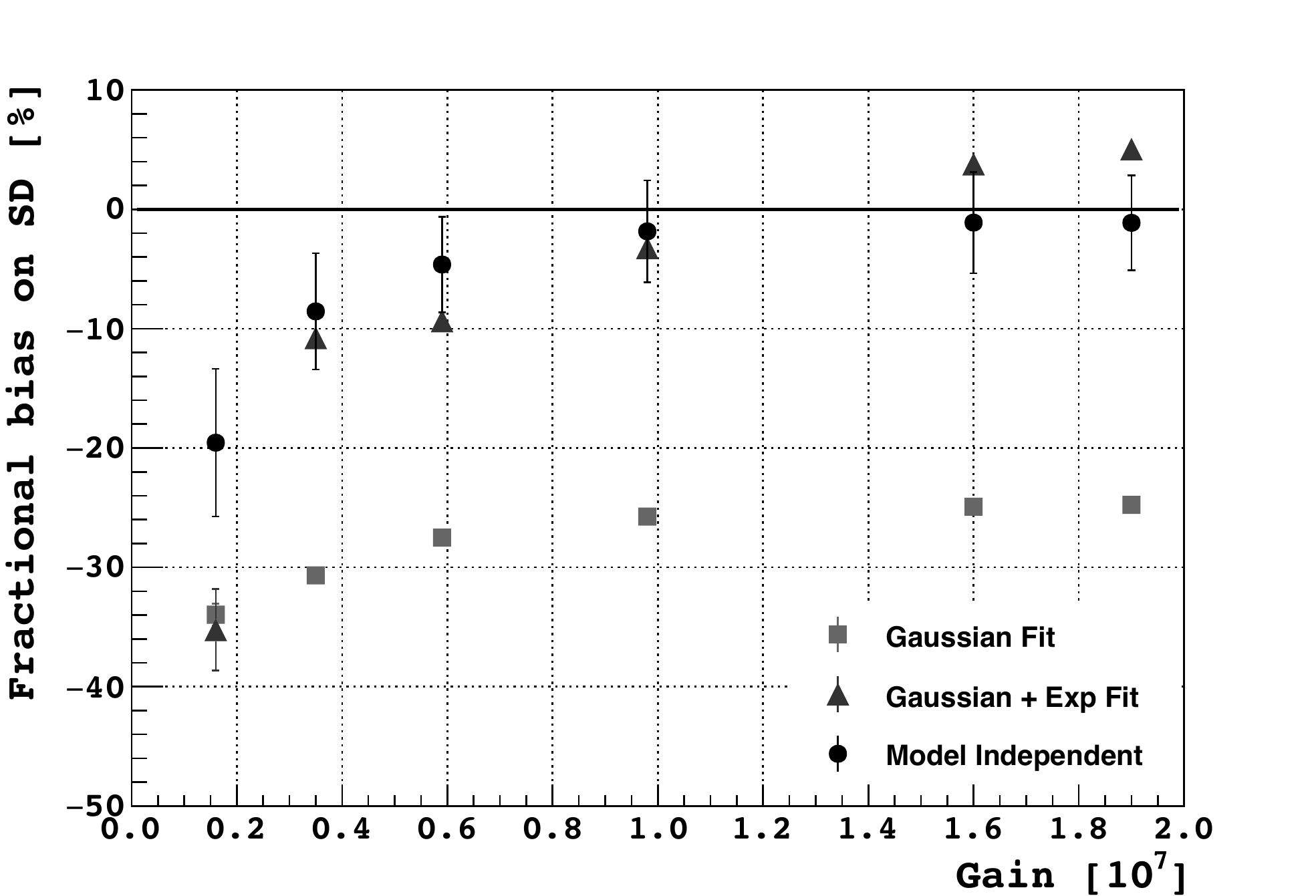}
\fi
\caption{Comparison of the estimated single photoelectron mean and standard deviation to more conventional fitting methods 
for a fixed occupancy (0.22 PE/trigger), with varying gain. The markers indicate the fractional bias while the error bars depict the fractional statistical uncertainty of the method.}
\label{fig:fit_comp_var_gain}
\end{center}
\end{figure*}
Finally, we compare the model independent method presented in this paper with more conventional fit methods. 
We have chosen two commonly used fitting methods which approximate the \spe~response as either a simple Gaussian 
distribution or a Gaussian + exponential \cite{Dossi2000623}. For both methods, the response to $p$ photoelectrons is assumed to be the $p-$times repeated 
convolution of the \spe~response, additionally convolved with a single Gaussian distribution representing the noise. 
Aside from specifying the shape of both the single photoelectron and noise response, 
these are the same underlying assumptions as the model independent method. 
We apply the fitting techniques to the simulated spectra that were generated with the empirically determined \spe~spectrum. Since the repeated convolution of a Gaussian + exponential 
is difficult to calculate exactly for more than 2 photoelectrons, we have restricted the comparison to a simulated data set with relatively low occupancy (\SI{0.22}{\pepertrigger}), to avoid biasing the fit results. 
Figure~\ref{fig:fit_comp_var_gain} shows the comparison of the bias in the estimated single photoelectron mean and standard deviation 
for the model independent method and the two fit methods. It can be seen that the model independent method provides a more accurate estimate 
than either of the fit methods at all gain values. The fit methods have comparatively smaller statistical uncertainties, though, as shown earlier, the statistical uncertainty of the model independent method can be reduced by either operating at higher occupancy or acquiring more statistics.
\section{Conclusions}
\label{sec:conclusions}
In this paper we have presented a simple new method to calibrate the single photoelectron response of photomultiplier tubes, taking into account the important contributions of under-amplified photoelectrons. Unlike conventional fitting methods, the proposed procedure determines the single photoelectron mean and variance statistically, without making any assumption about the underlying shape of the single photoelectron spectrum. It can therefore be used to calibrate \pmts~with different dynode structures, regardless of the fraction of under-amplified photoelectrons or the features of the single photoelectron charge spectrum. The method is shown to work well over a wide range (nearly two orders of magnitude) of light intensities and is therefore also suitable for the calibration of arrays of photomultipliers in large detectors, where uniform illumination is not possible. Following the description of the method, we have outlined the procedure to estimate the required parameters and their uncertainties, and applied the method to experimental data acquired with a Hamamatsu \pmtmodel~photomultiplier. Additionally we have used a Monte Carlo simulation with experimentally measured noise levels to study the results of the method as a function of the single photoelectron spectrum, photomultiplier gain, and light intensity. The method is found to estimate the single photoelectron mean and variance to better than 5\% for all single photoelectron spectra considered, at \pmt~gain values above $1\times10^7$. At lower gains, a small systematic bias is present due to the overlap of the under-amplified spectra with the noise; this bias can be reduced by temporarily raising the gain to accurately evaluate the occupancy or by the use of more sophisticated algorithms to distinguish between the signals and the background noise of the specific setup. Simulations using the SPE spectrum derived from our measurements of the R11410 \pmt~show that the method described here is more accurate than conventional fitting algorithms at all gain values. We note that our SPE spectrum is similar in shape to those obtained in other experimental setups with other \pmt~models \cite{Wright2010, Haas201106,ChirikovZorin2001310}, suggesting that our method has broad application.
\section{Acknowledgements}
\label{sec:acknowledgements}
The authors acknowledge helpful discussions with Jason Brodsky, Peter Meyers, and Alessandra Tonazzo during the development of this method. We are grateful to Stephen Pordes for his meticulous reading of the paper, which has significantly improved its clarity.  We also thank the staff of the Fermilab Particle Physics, Scientific, and Core Computing Divisions for their support. This work was supported in part by the Kavli Institute for Cosmological Physics at the University of Chicago through grant NSF PHY-1125897, NSF award PHY-1505581, and an endowment from the Kavli Foundation and its founder Fred Kavli, as well as the US DOE (Contract No. DEAC02-07CH11359).
\appendix
\section{}
\label{sec:stat_unc_occupancy}
From \eqn\eqref{eq:spe_mean_final}, the statistical uncertainty on the single photoelectron mean can be written as 
\begin{align}
\label{eq:stat_unc_basic}
\var{\sampmean{\psi}} &\approxim \frac{\var{\sampmean{\totalrand}} + \var{\sampmean{\bkgdrand}} +
\meansq{\sperand}\cdot \var{\estim{\occup}}}{\meansq{\estim{\occup}}}
\end{align}
where \sampmean{X} denotes the estimated mean, $i.e.$ the estimate of the mean of the distribution $X$ from the finite data sample taken and for convenience we have defined $\estim{\occup} \equiv \sampmean{\perand}$. We have ignored the smaller correlation terms.

Since the estimated mean of the total charge distribution, \sampmean{T}, is evaluated by calculating the arithmetic mean of the sampled distribution, 
the variance of the estimate is simply
\begin{align}
\var{\sampmean{\totalrand}} = \frac{\var{\totalrand}}{N_{L}}
\end{align}
where $N_{L}$~is the number of sample triggers in the laser data. Using \eqn\eqref{eq:spe_var_final}, one can write
\begin{align}
\label{eq:stat_unc_total}
\var{\sampmean{\totalrand}} = \frac{\mean{\perand}\cdot(\meansq{\sperand}+\var{\sperand}) + \var{\bkgdrand}}{N_{L}}
\end{align}

Similarly to \sampmean{\totalrand} and considering that $N_{B}$~is the number of sample triggers in the blank data, the variance of the estimate \sampmean{\bkgdrand} is
\begin{align}
\label{eq:stat_unc_background}
\var{\sampmean{\bkgdrand}} = \frac{\var{\bkgdrand}}{N_{B}}
\end{align}

The mean and variance of the estimate of the occupancy, $\estim{\occup} \equiv -\ln{(A_{T}/f\nsamps)}$, (excluding any bias from \nonzerope~triggers 
leaking below the amplitude cut) can be evaluated by first calculating the mean and variance of \nlaseramp, which is the number of \zerope~triggers 
that fall below the \thresholdcut. If the position of the \thresholdcut~cut was fixed, then \nlaseramp~would follow a simple Binomial distribution. However,
the position of the \thresholdcut~cut is chosen based on the randomly sampled background distribution and is therefore a random variable itself.
Thus the statistical uncertainty on \nlaseramp~depends on the fluctuations of two random processes:
\begin{enumerate}
\item Fluctuations in defining the position of the \thresholdcut~cut based on the background events.
\item Fluctuations in the number of laser events falling below the threshold.
\end{enumerate}

The fraction of the background distribution \bkgddist~that falls below a given charge value $q$ is simply the value of the cumulative background distribution, defined as \cumuldist{\bkgdrand}{q}. If the charge $q$ is itself sampled from the background distribution, then \cumuldist{\bkgdrand}{q} is also a random variable that follows a uniform distribution. The position of the \thresholdcut~cut $q_{t}$~is chosen to be the charge value of the $j = f\cdot(\nsampsbkgd+1)^{th}$ \textit{ordered} sample of the \nsampsbkgd~acquired background samples. Since the cumulative distribution is strictly increasing, it preserves order statistics and the fraction of the background distribution that falls below $q_t$ also corresponds to the value of the $j^{th}$ ordered sample of the uniform distribution \cumuldist{\bkgdrand}{q}. The $j^{th}$ order statistic of a uniform distribution is a beta random variable, and therefore $\phi_{t}\defined \cumuldist{\bkgdrand}{q_{t}}$ follows a beta distribution \betadist$(\alpha, \beta)$ with $\alpha \defined j,$ 
and $\beta \defined N_{B}-j+1\defined(1-f)\cdot(N_{B}+1)$.
\begin{align}
\betadist(\phi_{t}|j, N_{B}-j+1) &=  \frac{N_{B}!}{(j-1)!(N_{B}-j)!} \phi_{t}^{j-1}\phi_{t}^{N_{B}-j}
\end{align}
We can therefore analytically calculate the mean and variance of the distribution of $\phi_{t}$
\begin{align}
\mean{\phi_{t}} &= \frac{\alpha}{\alpha+\beta} 
= \frac{j}{N_{B}+1} 
= f\\
\var{\phi_{t}} &= \frac{\alpha\beta}{(\alpha+\beta)^2(\alpha+\beta+1)}  \nonumber\\
 &= \frac{f(1-f)}{N_{B}+2} 
\end{align}
As expected, the distribution of $\phi_{t}$ is unbiased with the mean equal to $f$ and its variance goes 
to 0 for large $N_{B}$.

Now, in the \laserdata~data set, we define the number of triggers that fall below $q_{t}$ to be \nlaseramp. 
\nlaseramp~does not follow a simple Binomial distribution, 
but instead follows a Binomial distribution where for each set of $N_{B}$~\blankdata~samples, 
the threshold $q_{t}$~varies according to the Beta distribution given above, 
and hence the probability of a zero-pe \laserdata~sample falling below $q_{t}$~varies. 
\begin{align}
\nlaseramp &\sim \binomdist(N_{L}, \pedistzero \phi_{t})\\
\phi_{t} &\sim \betadist(\alpha, \beta)
\end{align}
Note that for $\pedistzero = 1$, the distribution of \nlaseramp~would reduce to a standard beta-binomial distribution. One can compute the first two central moments using the following relations for mixture distributions:
\begin{flalign}
\mean{\nlaseramp} &=\displaystyle{\sum_{m=0}^{\infty}} m\displaystyle{\int_{0}^{1}} \binomdist(m | N_{L}, \pedistzero\zeropefrac) \cdot \betadist(\phi_{t} | \alpha, \beta) d\phi_t \nonumber\\
&= N_{L}\pedistzero f\\
\var{\nlaseramp}&=\displaystyle{\sum_{m=0}^{\infty}} m^2\displaystyle{\int_{0}^{1}} \binomdist(m | N_{L}, \pedistzero\zeropefrac) \cdot \betadist(\phi_{t} | \alpha, \beta) d\phi_{t} \nonumber \\
&~~~~- \meansq{\nlaseramp}\nonumber\\
&=  \textstyle{N_{L}\pedistzero f\left[(1-\pedistzero f) + \left(\frac{N_{L}-1}{N_{B}+2}\right)\pedistzero(1-f)\right]}\nonumber\\
&\approxim  \textstyle{N_{L}\pedistzero f\left[(1-\pedistzero f) + \pedistzero(1-f)\right]}
\end{flalign}
where in the last step we have used $N_{L}=N_{B}\defined N\gg1$ which we will assume from now on.

We can then write the mean and variance of the occupancy $\estim{\occup}\defined-\ln{(A_{T}/f N)}$ as
\begin{align}
\label{eq:stat_unc_occupancy_mean}
\mean{\estim{\occup}} &\approxim  -\loge{\pedistzero} = \lambda\\
\label{eq:stat_unc_occupancy_var}
\var{\estim{\occup}} &\approxim \frac{\left(\pedistzero^{-1}+1-2f\right)}{f N}\\
&=  \frac{\left(e^{\occup}+ 1 - 2\zeropefrac\right)}{f N}
\end{align}
Combining the individual statistical uncertainties from \eqns\eqref{eq:stat_unc_total}, \eqref{eq:stat_unc_background},  
\eqref{eq:stat_unc_occupancy_mean}, and~\eqref{eq:stat_unc_occupancy_var} into \eqn\eqref{eq:stat_unc_basic}, we get
\begin{align}
\var{\sampmean{\sperand}} \approxim&~\frac{\occup(\meansq{\sperand}+\var{\sperand}) + 2\var{\bkgdrand}}{\nsamps \occup^2} \notag\\
&+ \frac{\meansq{\sperand}\left(e^{\occup}+ 1 - 2\zeropefrac\right)}{\zeropefrac \nsamps \occup^2}
\end{align}

This can also be expressed in terms of the first two moments from the blank and laser distributions 
(\mean{\totalrand}, \mean{\bkgdrand}, \var{\totalrand}, \var{\bkgdrand})
\begin{align}
\var{\sampmean{\sperand}} \approxim&~\frac{\var{\totalrand} + \var{\bkgdrand}}{\nsamps \occup^2} \notag\\
&+ \frac{(\mean{\totalrand}-\mean{\bkgdrand})^2\left(e^{\occup}+ 1 - 2\zeropefrac\right)}{f \nsamps \occup^4}
\end{align}
\section{}
\label{sec:bias}
The presence of laser-induced photoelectron signals below the threshold cut can bias the estimated occupancy and 
consequently the estimated single photoelectron mean. The number of these \nonzerope~triggers, $l$, 
leaking below the \thresholdcut~cut can be reasonably expected to be proportional to the number of events that produce exactly 
one photoelectron, since the probability of two or more photoelectrons producing a combined signal that falls 
below the \thresholdcut~cut should be negligible.  One can then write,  \leakmean, the mean number of leakage events as
\begin{align}
\leakmean &= \speleakfrac \cdot \nsamps \cdot  \pedistone  \notag\\
&=  \speleakfrac \cdot \nsamps \cdot\occup \cdot \pedistzero 
\end{align}
where \speleakfrac~is the fraction of triggers with exactly one laser-induced photoelectron, 
whose total charge falls below the \thresholdcut~cut.  

Triggers with laser-induced single photoelectrons that fall below the \thresholdcut~cut lead to an overestimate of \nlaseramp, the number of \zerope~triggers below the cut  
\begin{align*}
\mean{\nlaseramp} &= \nsamps\pedistzero\zeropefrac + \leakmean \\
&= \nsamps\pedistzero\zeropefrac\left(1 + \frac{\speleakfrac}{\zeropefrac}\occup\right)
\end{align*}
This translates into biased estimates of the occupancy \biasestim{\occup} and the single photoelectron mean \biassampmean{\sperand}
\begin{align}
\mean{\biasestim{\occup}} &= \mean{-\loge{\nlaseramp/\zeropefrac\nsamps}}\notag\\
&\approxim \occup - \loge{1+\frac{\speleakfrac}{\zeropefrac}\cdot \lambda}\notag\\
&\approxim \occup \cdot \left(1 - \frac{\speleakfrac}{\zeropefrac}\right)\\
\mean{\biassampmean{\sperand}} &\approxim \mean{\sperand}\cdot \left(1 + \frac{\speleakfrac}{\zeropefrac}\right)
\end{align}
where we have assumed that $\speleakfrac/\zeropefrac \ll 1$.
\section*{References}
\bibliography{bibliography.bib}

\end{document}